\documentclass[12pt]{article}
\usepackage{epsf}
\usepackage{amsmath,amssymb}

\usepackage[dvips]{graphicx}

\usepackage{comment}

\begin{document}

\newcommand{\lsim}{\stackrel{<}{_\sim}}
\newcommand{\gsim}{\stackrel{>}{_\sim}}
\newcommand{\GI}{${\rm I}$}
\newcommand{\GII}{${\rm I\hspace{-.1em}I}$}
\newcommand{\GIII}{${\rm I\hspace{-.1em}I\hspace{-.1em}I}$}

\newcommand{\rem}[1]{{$\spadesuit$\bf #1$\spadesuit$}}

\renewcommand{\thefootnote}{\fnsymbol{footnote}}
\setcounter{footnote}{0}

\begin{titlepage}

\def\thefootnote{\fnsymbol{footnote}}

\begin{center}
\hfill KEK-TH-1904\\
\hfill MISC-2016-04\\
\vskip .75in

{\LARGE \bf 



MSSM without a free parameter

}

\vskip 0.5in

{\large 
Ryuichiro Kitano$^{1,2}$,
Ryuji Motono$^{1,2}$
and Minoru Nagai$^3$
}

\vskip 0.25in

$^1$
{\em Institute of Particle and Nuclear Studies,\\
High Energy Accelerator Research Organization (KEK)\\
Tsukuba 305-0801, Japan}

\vskip 0.1in

$^2$
{\em The Graduate University for Advanced Studies (Sokendai)\\
Tsukuba 305-0801, Japan}

\vskip 0.1in

$^3$
{\em Maskawa Institute for Science and Culture, Kyoto Sangyo University, \\
Kyoto 603-8555, Japan
}

\end{center}
\vskip .5in

\begin{abstract}
It is often argued that the minimal supersymmetric standard model has
 $O(100)$ free parameters and the generic parameter region is already
 excluded by the null observation of the flavor and CP violating
 processes as well as the constraints from the LHC experiments.
%
%
This situation naturally leads us to consider the case where all the
 dangerous soft supersymmetry breaking terms such as the scalar masses
 and scalar couplings are absent, while only the unified gaugino mass
 term and the $\mu$ term are non-vanishing at the grand unification scale.
%
%
We revisit this simple situation taking into account the observed Higgs
 boson mass, 125~GeV. Since the gaugino mass and the $\mu$ term are
 fixed in order to explain the Higgs boson and the $Z$ boson masses,
 there is no free parameter left in this scenario.
We find that there are three independent parameter sets exist including
 ones which have not been discussed in the literature. We also find
 that the abundance of the dark matter can be explained by relic gravitinos
 which are non-thermally produced as decay products of the SUSY
 particles while satisfying constraints from Big Bang Nucleosynthesis.
 We discuss the effects of the gravity mediation which generically give
 contribution to the soft terms of the order of the gravitino mass. It
 turns out that newly found parameter set is preferable to explain the
 Higgs boson mass as well as the gravitino dark matter while satisfying
 the constraints from the electric dipole moments of the electron and
 the nucleon.
\end{abstract}

\end{titlepage}

\renewcommand{\thepage}{\arabic{page}}
\setcounter{page}{1}
\renewcommand{\thefootnote}{\#\arabic{footnote}}
\setcounter{footnote}{0}


\section{Introduction} \label{sec:introduction}
The Higgs boson mass, 125~GeV, suggests that the physics behind the
electroweak symmetry breaking is weakly coupled, but it is not quite as
light as the predictions of TeV scale supersymmetry.
%
%
In order to explain the Higgs boson mass in the minimal supersymmetric
standard model (MSSM), the superpartner masses especially the scalar top
quarks needs to be above ${\cal O}(10)$~TeV in a generic region of the
parameter space \cite{Ibe:2012qu,Draper:2013oza,Hahn:2014qla,14074081}.


In light of this situation, models with ${\cal O}$(PeV) scale rather
than ${\cal O}(10)$~TeV supersymmetry have been discussed quite
extensively \cite{Ibe:2011aa,Ibe:2012hu,Arvanitaki:2012ps,ArkaniHamed:2012gw}. (For earlier studies, see \cite{9810442,ArkaniHamed:2004fb,Wells:2004di,Hall:2011jd}.) The reason for this jump from TeV to PeV is based on the
constraints from the flavor and CP violating processes \cite{Moroi:2013sfa,McKeen:2013dma,Altmannshofer:2013lfa}. The flavor and
CP constraints can be (almost) avoided by raising the SUSY scale to PeV,
while the dark matter of the Universe can be explained by the thermal
relic of the light gauginos, in particular the Wino \cite{0610249}, whose masses can be
suppressed by a one-loop factor (and thus TeV scale) as in the anomaly
mediation scenario \cite{9810442,9810155}.


On the other hand, there is another natural framework with ${\cal
O}(10)$~TeV SUSY. Since we need some mechanism to suppress large flavor
and CP violations, the gravity-mediated contributions which generically
break flavor and CP should be suppressed, i.e., the gravitino should be
much lighter than other SUSY particles. In this case, the gravitino can
be a good candidate of the dark matter; its abundance may be
explained by the decay of other SUSY particles which are thermally
produced \cite{Feng:2003xh,Feng:2003uy}.

In this paper, we discuss the scenario where all SUSY breaking
parameters except gaugino mass are set to be zero at the cut-off scale,
and thus the severe flavor and CP constraints can be alleviated.  
Such a set-up has been studied as the low energy effective theory of the
gaugino mediation scenario \cite{Kaplan:1999ac,Chacko:1999mi} or the no-scale supergravity Lagrangian \cite{Ellis:1983sf,noscale}.
The right-handed stau tends to be the Next-to-Lightest SUSY Particle
(NLSP) and its decay into the gravitino may explain the observed dark
matter abundance. However it is non-trivial whether the Higgs boson
mass and dark matter abundance can be explained simultaneously in this
simple set-up.  The large stau mass is required to avoid the severe
constraints from Big Bang Nucleosynthesis (BBN) while stop masses are
bounded from above according to the value of $\tan\beta$ to explain the
observed Higgs boson mass.

We discuss the running behavior of the Higgs $B$ term carefully and find
that a new parameter set with relatively small $\tan\beta$ appears
for a large SUSY breaking scale even with vanishing Higgs $B$ term at
the cut-off scale.  This helps to solve the above-mentioned tension;
small $\tan\beta$ enlarges the right-handed stau mass and also relax the
upper bound on stop masses.  The small gravity-mediated contributions can
become the sources of flavor and CP violation in the model.  However we
find that, thanks to the small value of $\tan\beta$, the predicted
electron Electric Dipole Moment (EDM) is marginal to the present
experimental bound and should be checked in the near future experiments.

This paper is organized as follows.  In Section~\ref{sec:model} we
explain our model to solve the SUSY flavor and CP problems.  The
renormalization group running of the Higgs $B$ term is examined
carefully and we identify the parameter regions with the correct
electroweak symmetry breaking (EWSB) minimum.  The spectrum of SUSY
particles and the lightest Higgs boson are also presented here.
We discuss the implications on the gravitino dark matter in
Section~\ref{sec:gravitino} and predictions of the electron and nucleon
EDMs in Section~\ref{sec:edms}. In Section~\ref{sec:thermal}, we discuss the thermal component
of the gravitino relic abundance. We summarize our results in Section~\ref{sec:summary}.

\section{CP- and Flavor-Safe Minimal SUSY Model} \label{sec:model}
In the minimal supersymmetric standard model (MSSM), independent CP-phases are expressed in the combination of 
$A$ term, $B$ term, $\mu$ term and gaugino masses, $M_i$.
\begin{eqnarray}
  \phi_{\mu, i} = {\rm arg}\left( M_i \mu (B\mu)^* \right), \quad
  \phi_{{A_f}, i} = {\rm arg}\left( M_i  A_f^* \right) \quad (f=u,d,e)
\end{eqnarray}
Usually we take the basis with real $B \mu$ by the appropriate redefinitions of fields so that
the vacuum expectation values (VEVs) of the two Higgs doublets become real. 
Flavor and CP violations come from the off-diagonal terms of the sfermion mass terms in the SCKM basis, 
\begin{eqnarray}
  \left( m^2_{\tilde f} \right)_{ij}  \quad ( i\ne j)~.
\end{eqnarray}
If the SUSY scale is below 100 {\rm TeV}, random values of these parameters predict
detectable FCNC and CP-violating phenomena.

We assume that all of $A$ terms, $B$ terms and sfermion soft masses vanish at some scales,
\begin{eqnarray}
    A_{u,d,e} = B = m^2_{{\tilde q},{\tilde u},{\tilde d},{\tilde l},{\tilde e}}=0~,
    \label{eq:safe}
\end{eqnarray}
At the low energy scale, non-vanishing $A$ and $B$ terms are generated by the radiative corrections 
through the gauge interactions.
Since these contributions are proportional to gaugino masses, there appear no CP phases 
at the low energy scale as long as phases of three gaugino mass parameters are aligned. 
The off-diagonal elements of sfermion mass are also generated only radiatively through the CKM matrix,
and the flavor constraints can become rather weak. 
The remaining free parameters are Higgs soft masses and gaugino masses, 
\begin{eqnarray}
 m^2_{H_u}, \quad m^2_{H_d}, \quad M_{1,2,3}~.
\end{eqnarray}
We consider the minimal situation where the SUSY breaking is directly mediated only 
to the gauge sectors by the physics of grand unification theories (GUTs). 
Then, we further impose following conditions at the GUT scale, $M_{\rm G}$,
\begin{eqnarray}
  m^2_{H_u}=m^2_{H_d}=0, \quad M_1=M_2=M_3=M_{1/2}.
  \label{eq:safe2}
\end{eqnarray}
In this way, one can consider a very predictive framework where we have
only one SUSY breaking parameter, $M_{1/2}$, and one supersymmetric
parameter $\mu$. Note, the size of the SUSY breaking is naively
estimated as $M_{1/2} \simeq {\cal O} (F/M_{\rm G})$ and thus the
gravitino becomes the lightest SUSY particle with mass $m_{3/2}=F /
(\sqrt{3}M_{\rm pl}) \simeq 10^{-(2-3)} M_{1/2}$.  The small gravitino
mass is also favored to suppress the possibly dangerous gravity-mediated
contributions, which are the main sources of flavor- and CP-violations
in our scenario.

In this model, the ratio of the VEVs, $\tan \beta = \langle H_u \rangle
/ \langle H_d \rangle$ is not a free parameter and is determined by the
condition of the EWSB.
At the SUSY scale, ${\cal O}(10)$~TeV, the following conditions should
be imposed:
%
\begin{eqnarray}
 \frac{m_Z^2}{2}  \!\!\!\! &=&\!\!\! 
   -  |\mu|^2 +  \frac{ m^2_{H_u} +\Sigma_u}{\cot^2\beta-1} 
   - \frac{m^2_{H_d}  +\Sigma_d}{1-\tan^2\beta} ~, \label {eq:EWSB1} \\
 \sin 2\beta  \!\!\!\! &=& \!\!\! - \frac{B \mu} {2 |\mu|^2+ m^2_{H_u} + m^2_{H_d} +\Sigma_u +\Sigma_d}. \label{eq:EWSB2}
\end{eqnarray}
where $\Sigma_{u,d}$ includes the tadpole contributions originated from the one-loop corrections
\footnote{Here, we use the 1-loop effective potential to determine only $\mu$, $\beta$ and masses of heavy Higgs bosons.
On the other hand  the lightest Higgs boson mass is calculated by the effective field theory approach following \cite{14074081,11086077,13073536} 
since we consider a relatively high SUSY scale to explain the observed Higgs boson mass.
}. 
For a given choice of $M_{1/2}$, these two constraints fix the $\mu$
parameter as well as the value of $\tan \beta$ by requiring that the
$B$ parameter vanishes at the unification scale.
We choose $0<\beta<\pi/2$ to obtain positive VEVs of two Higgs doublets.
In this convention, the sign of the $B$ term determines that of the
$\mu$ term.  The low energy value of the $B$ parameter is evaluated by
the following renormalization group (RG) equation \footnote{ In the
actual calculation of the SUSY spectrum we use two-loop renormalization
group equations above the SUSY scale.  },
\begin{eqnarray}
  16 \pi^2 \frac{d B}{d \log{\mu}} \simeq - 3 g_2^2 M_2 - g_1^2 M_1 + y_\tau^2 A_\tau + 3 y_b^2 A_b + 3y_t^2 A_t~.
\end{eqnarray}
Here, the bottom Yukawa coupling receives sizable threshold corrections for large $\tan\beta$, 
and its value is sensitive to the size of $\tan\beta$ and the sign of $\mu$, 
\begin{eqnarray}
 y_b(\mu_{\rm SUSY}) \simeq \frac{g_2 m_b}{\sqrt{2} m_W} \frac{\tan\beta}{1+\varepsilon_{b} \tan\beta}, \quad 
 \varepsilon_{b} \approx \frac{\alpha_s}{3\pi} \frac{\mu M_3}{m_{\tilde q}^2}~.
\end{eqnarray}

\begin{figure}[tp]
  \centerline{
  \epsfxsize=0.45\textwidth \epsfbox{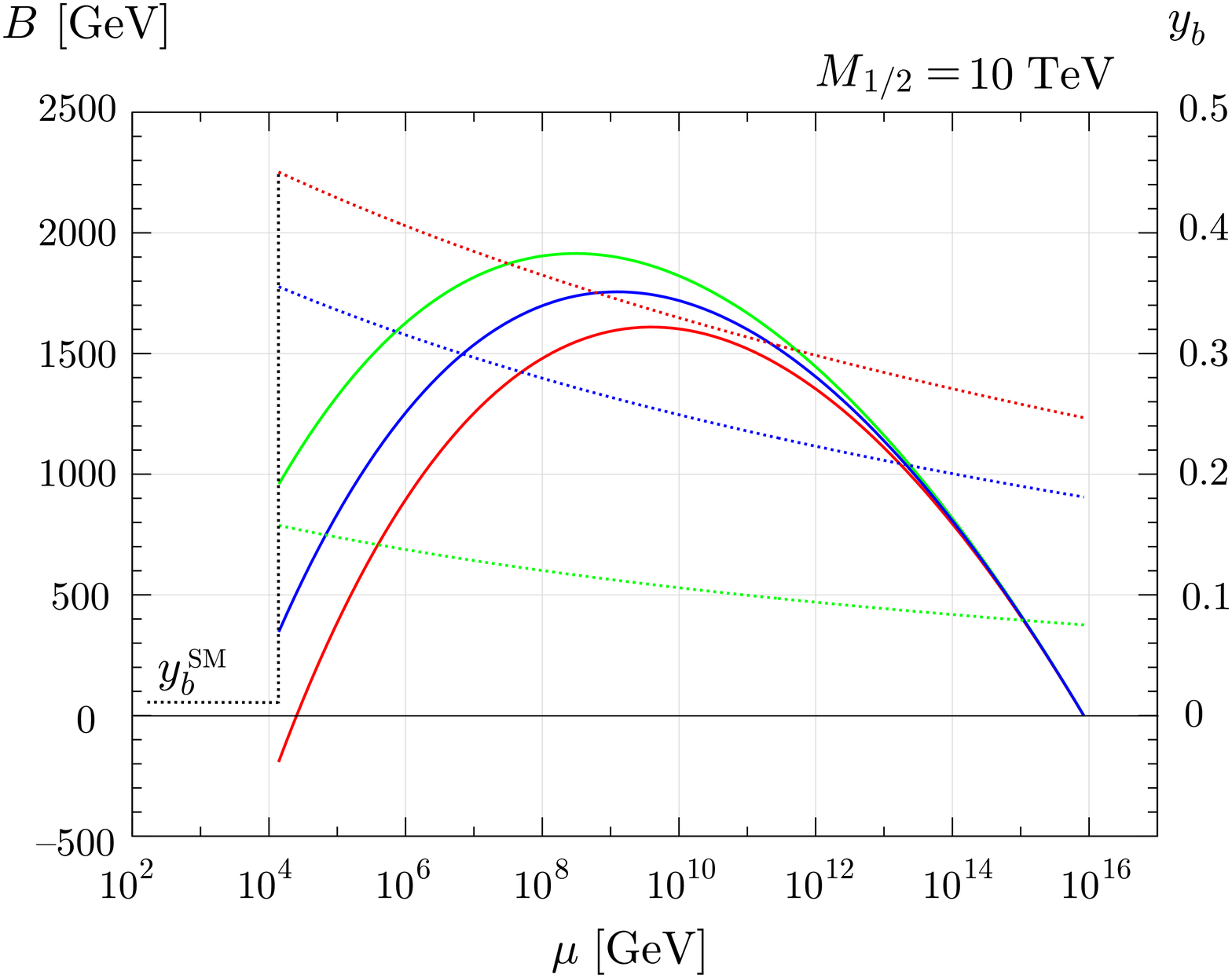}
  \epsfxsize=0.45\textwidth \epsfbox{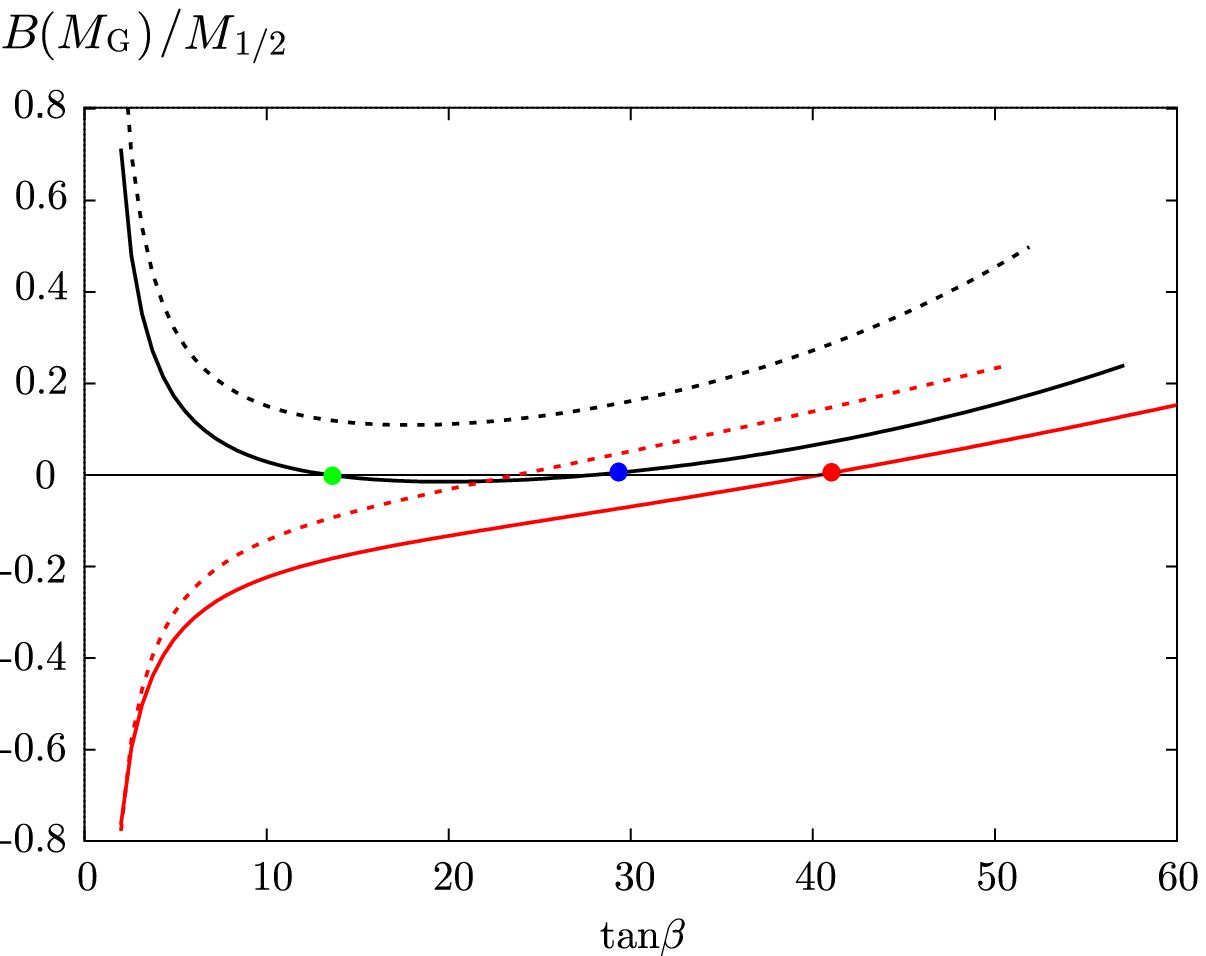}
  }
  \caption{ ({\it Left}):
  The running behaviors of the Higgs $B$ parameter (solid lines with left y-axis) and 
  the bottom Yukawa couplings (dotted lines with right y-axis) for $M_{1/2}=10~{\rm TeV}$. 
  The each colored line corresponds to the solution with (\GI) $\mu > 0$ (red), (\GII) $\mu< 0$ with 
  large $\tan\beta$ (blue) and (\GIII) $\mu<0$ with small $\tan\beta$ (green). 
  ({\it Right}):
  The values of the B parameter at the GUT scale as a function of $\tan\beta$. 
  The red (black) lines corresponds to the solution with $\mu>0$ ($\mu<0$). 
  The B parameters are normalized by $M_{1/2}$ and dotted (solid) lines correspond 
  to the case for $M_{1/2} = 1\ (10) ~{\rm TeV}$. 
  The red, blue and green dots correspond to the three solutions \GI, \GII~and \GIII~
  for $M_{1/2}=10~{\rm TeV}$.
  }
  \label{fig:B}
\end{figure}

The typical running behaviors of the $B$ parameter and the bottom Yukawa coupling are presented 
in Figure \ref{fig:B} (left).
At a high energy scale, the $B$ parameter is increased by the gauge interactions 
as the renormalization scale $\mu$ goes down. 
Since the values of $A$ terms are also enhanced at the low energy scale, 
the $B$ parameter turns to be decreased by the Yukawa interactions. 
We find three solutions to satisfy the EWSB conditions, 
\begin{enumerate}
\setlength{\leftskip}{0.5 cm}
\item[(\GI)]  { $\mu>0$, $B<0$ and large $\tan\beta$}
\item[(\GII)]  {$\mu<0$, $B>0$ and large $\tan\beta$ }
\item[(\GIII)]  {$\mu<0$, $B>0$ and small $\tan\beta$ }
\end{enumerate}
In the scenario \GI,  values of $y_b$ and $\tan \beta$ are large enough to drive the $B$ parameter negative. 
The large $\tan\beta$ implies a small absolute value of $B$ parameter according to Eq.~(\ref{eq:EWSB2}). 
For smaller but still large $y_b$, the $B$ parameter keeps positive even at the SUSY scale and its absolute value 
is small in the scenario \GII. For much smaller $y_b$, we obtain a large $B$ parameter, implying small $\tan\beta$ 
in the scenario \GIII. 
We note that the large $B$ parameter is preferable to suppress CP phases generated 
by the gravity-mediated contributions, which are estimated as $\phi_\mu \sim m_{3/2}/|B|$.

The solutions \GII~and \GIII~do not appear when the SUSY scale is low since the contributions 
from Yukawa couplings are much effective in the low energy region, always resulting in a negative $B$ parameter.
In Figure~\ref{fig:B} (right), we show the values of $B$ parameter at the GUT scale as functions of $\tan\beta$ for 
$M_{1/2}=1~{\rm TeV}$ and $10~{\rm TeV}$. 
For large $\tan\beta$, the $B$ parameter has to be large enough at the GUT scale to have small absolute value 
at the SUSY scale since it receives large negative contributions coming from the large bottom Yukawa coupling. 
Then, we always have a solution of $B(M_{\rm G})=0$ for $\mu>0$ (red lines) irrespective of $M_{1/2}$. 
On the other hand, the solutions with $\mu<0$ (black lines) appear only for large $M_{1/2}$ since 
the negative contributions of the Yukawa couplings are weakened for the high SUSY scale. 


This argument shows that the appearance of parameter regions with $\mu<0$ are sensitive to the precise 
sizes of Yukawa couplings and also to the GUT scale $M_{\rm G}$. 
We take the GUT scale as the scale with $g_1(M_{\rm G})=g_2(M_{\rm G})$, which becomes a little bit 
smaller for the higher SUSY scale.
The precise value of the top Yukawa couplings is also essential to calculate the Higgs boson mass 
in the SUSY model. 
Therefore, in the following analysis, we take a relatively large uncertainty for the top pole mass, 
$M_t = 173.3 \pm 2~{\rm GeV}$ compared to the result obtained by the LHC experiments \cite{Chatrchyan:2013xza,Abazov:2014dpa,Aaltonen:2014sea}, taking into account 
the possible difference between the measured mass parameter and the pole mass. 
And also, since we are focusing on high-scale SUSY models, we adapt the effective field theory approach to 
calculate the lightest Higgs boson mass. 
Concretely, we use 3-loop Standard Model (SM) RG equations to calculate the Yukawa couplings at the SUSY scale ($\sim M_{1/2}$) 
and to obtain Higgs quartic coupling at the SM scale ($\sim M_t$). 
Appropriate threshold corrections are included according to \cite{14074081,11086077,13073536} both at the EW scale and at the SUSY scale.
The SUSY spectrum is calculated by solving 2-loop RG equations
\footnote{
We compared our results with those obtained by a modified version of SOFTSUSY \cite{0104145}, 
which reproduces the gauge and Yukawa couplings derived by the effective field approach at the high energy scale, 
and we found that the difference of obtained SUSY masses are within 1 \%.
}.
Here, we stress again that our model now has only one parameter,
$M_{1/2}$, that can be determined uniquely for each solutions,
\GI, \GII~and \GIII~, to reproduce the observed Higgs boson mass,
125~GeV, up to the current experimental uncertainties in
the determinations of the Higgs boson mass and the top quark mass as
well as the theoretical uncertainties in the calculations of the mass
spectrum of the superparticles.
%

\begin{figure}[tp]
  \centerline{  
  \epsfxsize=0.45\textwidth \epsfbox{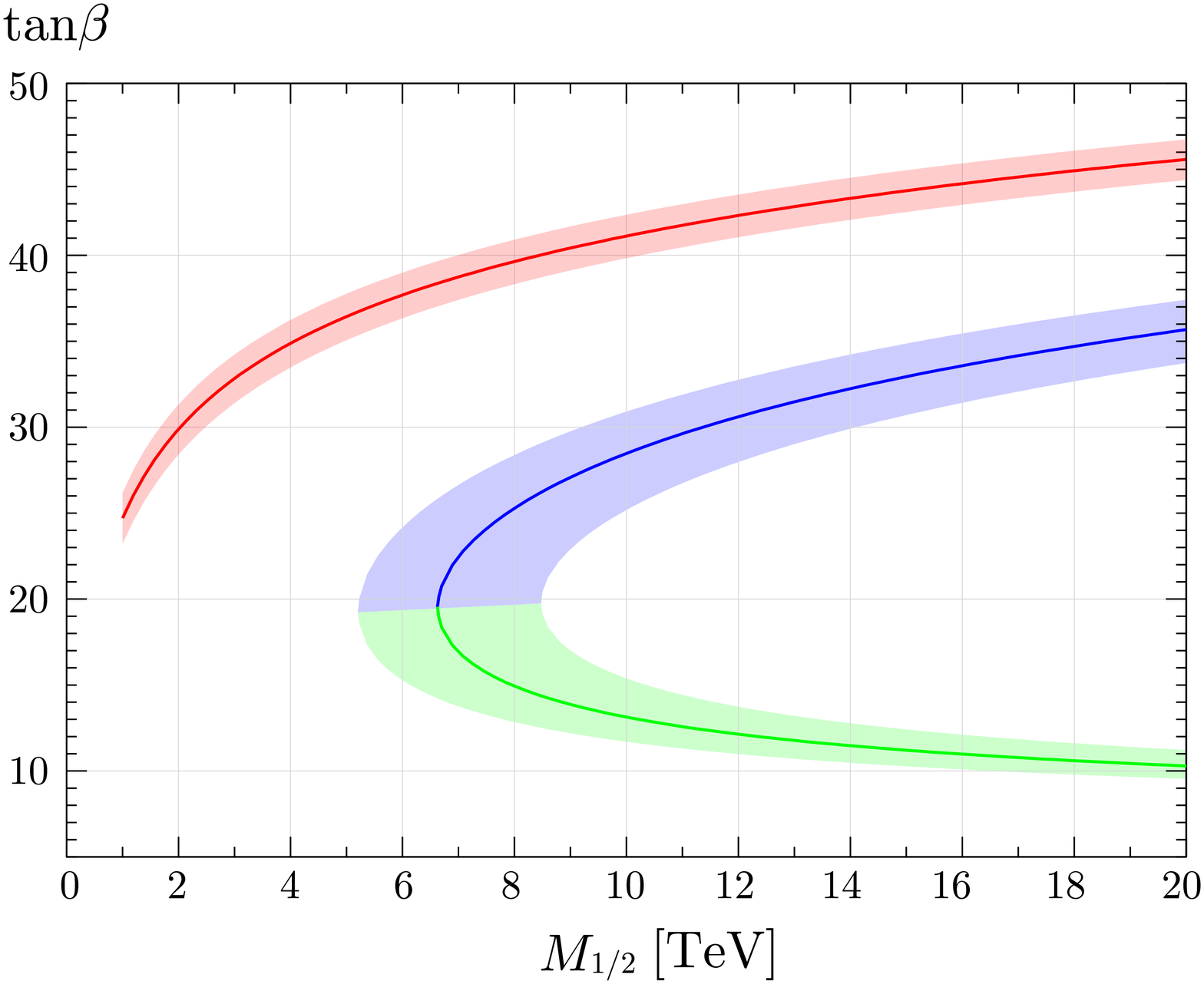} 
  \epsfxsize=0.45\textwidth \epsfbox{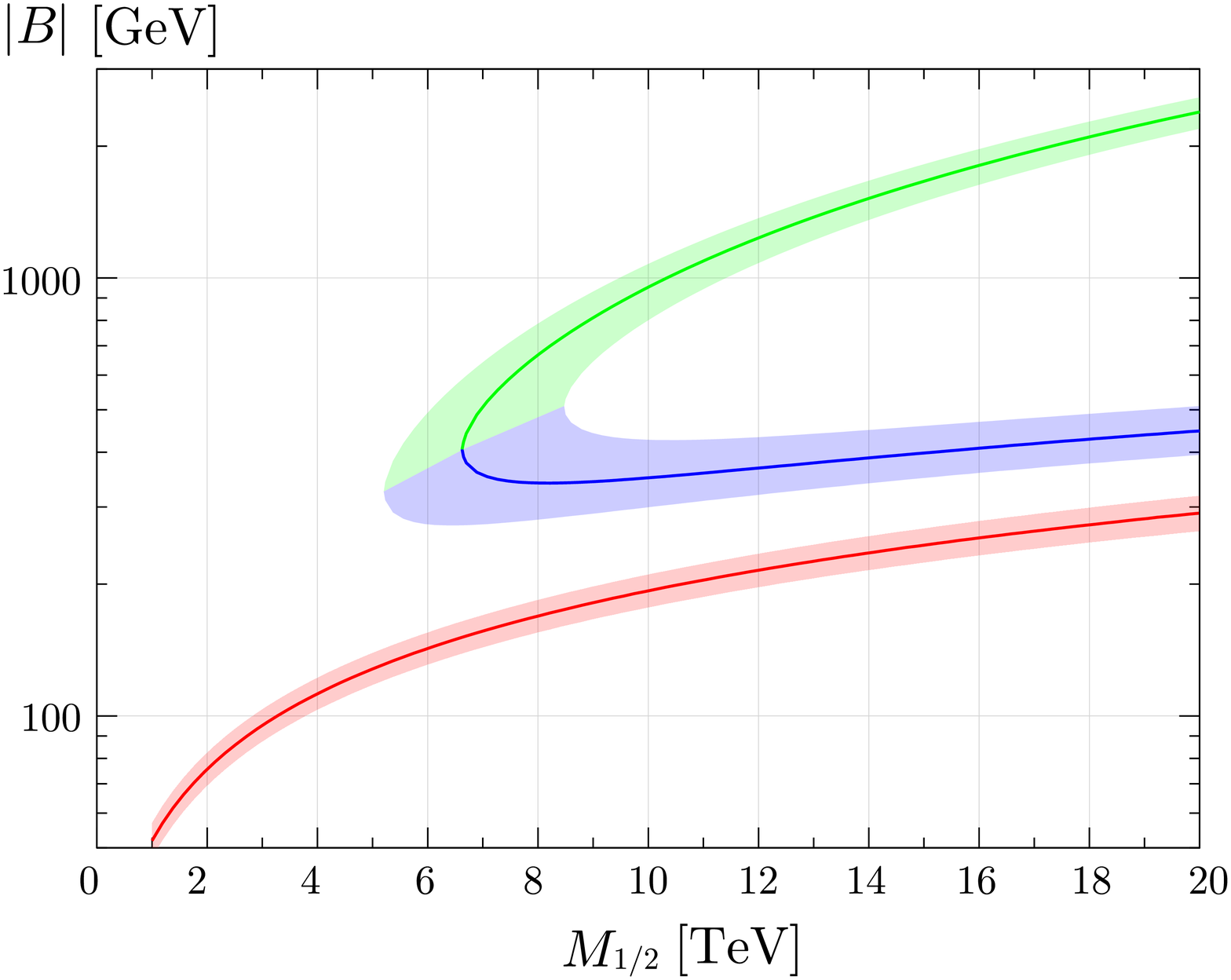} 
  }
  \caption{
  The values of $\tan\beta$ (left) and $|B|$ (right) as a function of $M_{1/2}$.
  The red, blue and green lines correspond to the scenario \GI, \GII~and \GIII~respectively. 
  The colored  bands show the uncertainty coming from the top mass, $M_t=173.3 \pm 2~{\rm GeV}$.
  }
  \label{fig:tanbeta}
\end{figure}

In Figure~\ref{fig:tanbeta} we show the values of $\tan\beta$ and $|B|$ at the SUSY scale for the each solution.
In the case of the solutions \GI~and \GII, $\tan\beta$ is a monotonically increasing function of $M_{1/2}$. This is because 
negative contributions from the top Yukawa coupling is weakened for large SUSY scale and the bottom Yukawa 
coupling has to be larger to obtain small $|B|$. 
On the other hand, $\tan\beta$ becomes smaller for large $M_{1/2}$ in the case of the solution \GIII, 
simply because we obtain larger $|B|/M_{1/2}$ for larger $M_{1/2}$ 
due to the smaller contributions from Yukawa couplings and it implies smaller $\tan\beta$ according to Eq.(\ref{eq:EWSB2}).
The obtained size of $|B|$ is about ${\cal O}(100)~{\rm GeV}$ which is much smaller than 
$M_{1/2} \sim {10}~{\rm TeV}$ because of large cancellation between gauge interactions and Yukawa interactions. 
However, $|B|$ could be large enough, $|B| \gsim 1~{\rm TeV}$, in the case of the solution \GIII~because negative contributions from Yukawa couplings are weakened thanks to 
the high SUSY scale and the low value of $\tan\beta$.

\begin{figure}[tp]
  \centerline{
  \epsfxsize=0.45\textwidth\epsfbox{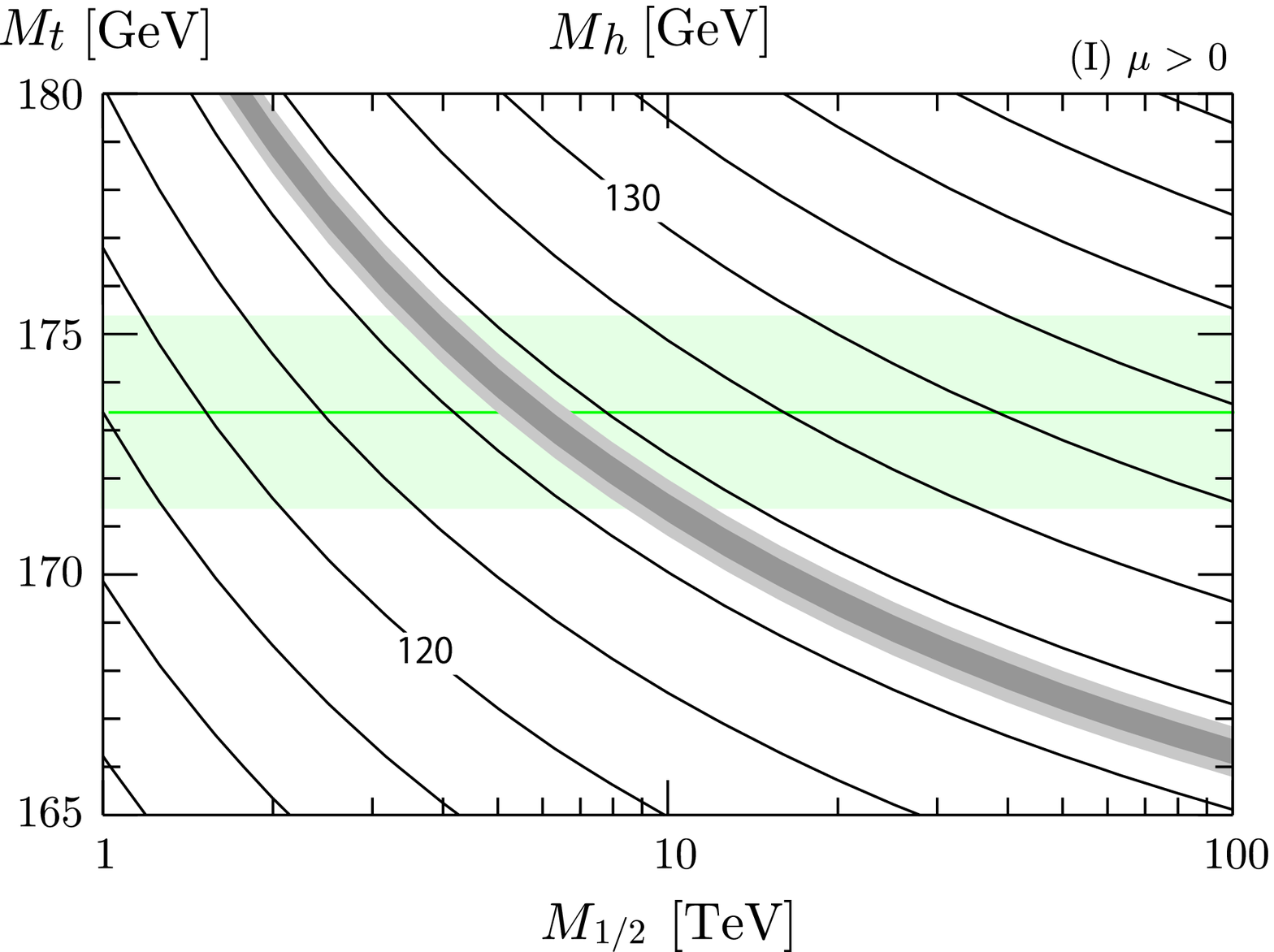}
  \epsfxsize=0.45\textwidth\epsfbox{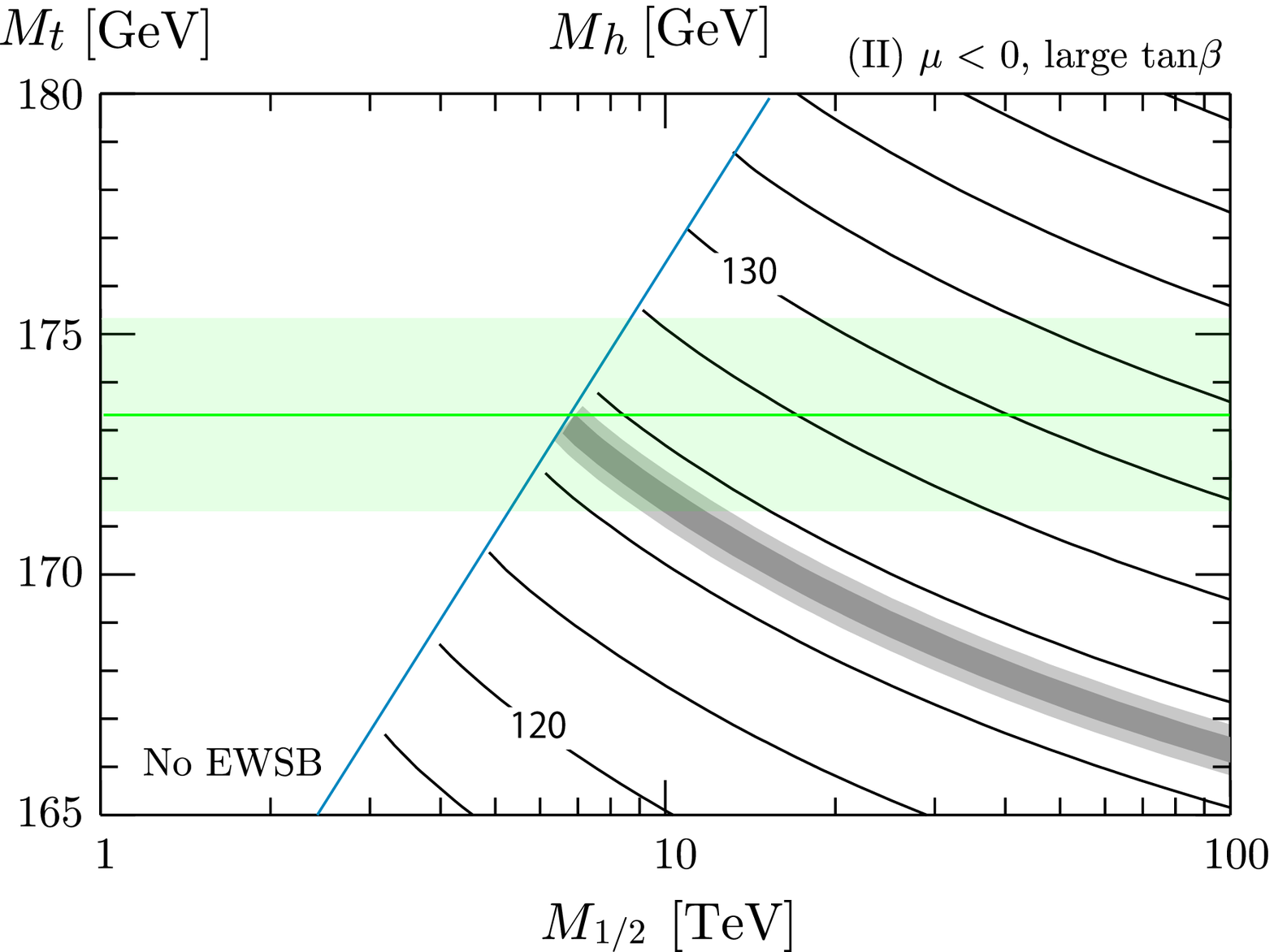}
  }
  \vspace{0.5cm}
  \centerline{
  \epsfxsize=0.45\textwidth\epsfbox{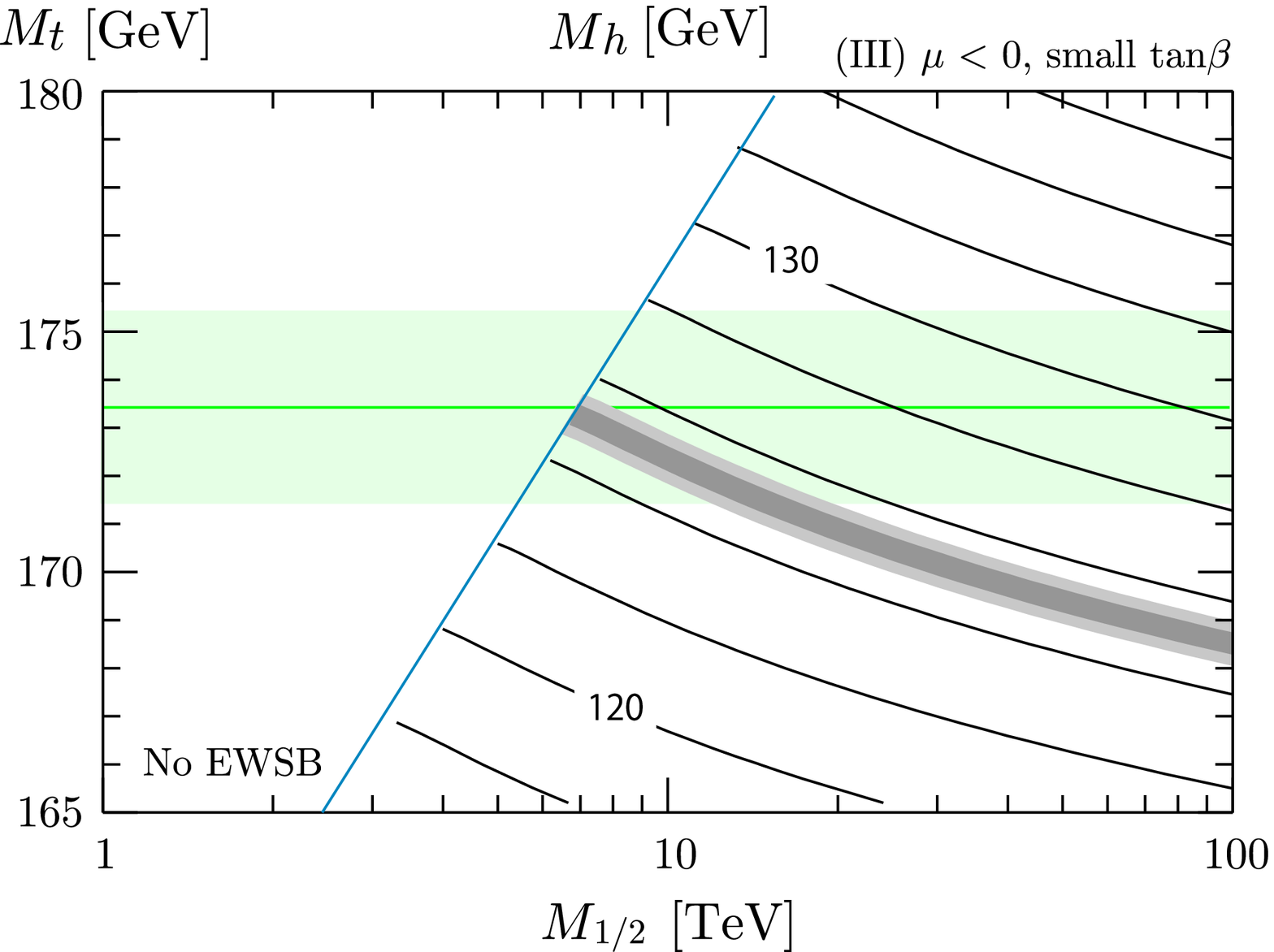}
  }
  \caption{ 
  Contour plots for the mass of the Higgs boson as functions of $M_{1/2}$ and $M_t$ 
  in the scenario \GI~(top left), \GII~(top right) and \GIII~(bottom). 
  Gray shaded regions are favored by the measurements of the Higgs boson mass by the LHC experiments.
  Green regions indicate the uncertainty of the top mass.
  }
  \label{fig:mHiggs}
\end{figure}

In Figure \ref{fig:mHiggs} we show the parameter regions which explain the correct 
Higgs boson mass, $m_h = 125.09 \pm 0.24~{\rm GeV}$ \cite{Aad:2015zh}, 
in the ($M_{1/2}$, $M_t$) plane for the scenario \GI~(top left), \GII~(top right) 
and \GIII~(bottom).
Due to the uncertainty of the top mass, 
$M_{1/2}$ is not determined uniquely and instead we obtain upper and lower bound on it. 
For the solution \GI, the observed Higgs boson can be explained in the range of 
$3.5~{\rm TeV}\lsim M_{1/2}\lsim 12~{\rm TeV}$ depending on the top mass. 
In the case of the solutions \GII~and \GIII, the SUSY scale have to be large enough to realize the EWSB, 
and also a little bit larger $M_{1/2}$ is required for the solution \GIII~since $\tan\beta$ is relatively small.
We find $6.5~{\rm TeV}\lsim M_{1/2}\lsim 13~{\rm TeV}$ for the solution \GII~and 
$6.5~{\rm TeV}\lsim M_{1/2}\lsim 20~{\rm TeV}$ for the solution \GIII.
We find that the top mass should be less than 174~{\rm GeV} in the scenario \GII~and \GIII. 
As the top mass becomes smaller, a larger $M_{1/2}$ is necessary to have enough radiative corrections 
to the Higgs mass. 
Especially the required $M_{1/2}$ is increased rapidly in the solution \GIII~since the tree-level contributions 
to the Higgs mass, which is proportional to $\cos 2\beta$, is also decreased for larger $M_{1/2}$.
This figure shows that the precise measurements of the masses of the top quark and the Higgs boson are
essential to confirm or exclude our model.

\begin{table}[tp]
  \begin{center}
  \begin{tabular}{|c|c|c|c|}
    \hline
      $\qquad \qquad \qquad$ & \GI $\ (\mu>0)$ & \GII $\ (\mu<0)$ & \GIII $\ (\mu<0)$ \\
    \hline
    \hline
    $M_t\ {\rm  [GeV]}$  & 173.3 & 172.0 & 172.0 \\
    \hline
    $M_{1/2}\ {\rm  [TeV]}$ & 5.0 & 10.0 & 15.0 \\
    \hline
    $\tan\beta$             & 36.5   & 30.3 & 10.5 \\
    $B \ {\rm [GeV]}$    & -127   & 312 & 1776 \\
    $M_h\ {\rm  [GeV]}$ & 124.5 &125.4 & 125.6 \\
    $M_A \ {\rm [TeV]} $ & 4.85   & 9.29 & 15.5 \\
    \hline
    $M_1 \ {\rm [TeV]} $ & 2.30   & 4.73 & 7.21 \\
    $M_2 \ {\rm [TeV]} $ & 4.05   & 8.19 & 12.4 \\
    $M_3 \ {\rm [TeV]} $ & 9.65   & 18.6 & 27.3 \\
    $\mu \ {\rm [TeV]}  $ & 4.64   & -8.38 & -12.1 \\
    \hline
    $m_{{\tilde e}_R} \ {\rm [TeV]}    $  & 1.80 & 3.58 & 5.35 \\
    $m_{{\tilde \tau}_R} \ {\rm [TeV]} $ & 1.39 & 3.17 & 5.27 \\
    $m_{{\tilde e}_L} \ {\rm [TeV]}     $ & 3.16 & 6.22 & 9.24 \\
    $m_{{\tilde \tau}_L} \ {\rm [TeV]} $ & 3.06 & 6.11 & 9.22 \\
    $m_{{\tilde u}_R} \ {\rm [TeV]}    $ & 8.08 & 15.3 & 22.3 \\
    $m_{{\tilde t}_R} \ {\rm [TeV]}     $ & 6.84 & 13.1 & 19.0 \\
    $m_{{\tilde d}_R} \ {\rm [TeV]}    $ & 8.02 & 15.2 & 22.1\\
    $m_{{\tilde b}_R} \ {\rm [TeV]}    $ & 7.64 & 14.6 & 22.0 \\
    $m_{{\tilde u}_L} \ {\rm [TeV]}     $ & 8.50 & 16.2 & 23.6 \\
    $m_{{\tilde t}_L} \ {\rm [TeV]}      $ & 7.76 & 14.9 & 22.1 \\
    \hline
    $A_u$  \ {\rm [TeV]}      & 9.22  & 17.3 & 25.0 \\    
    $A_t$   \ {\rm [TeV]}      & 7.33  & 14.0 & 20.7\\    
    $A_d$  \ {\rm [TeV]}      & 10.3  & 19.4 & 29.0 \\    
    $A_b$  \ {\rm [TeV]}      & 9.26  & 17.6 & 27.5 \\    
    $A_e$  \ {\rm [TeV]}      & 2.16  & 4.42 & 7.65 \\    
    $A_{\tau}$ \ {\rm [TeV]}  & 1.96  & 4.21 & 7.61 \\    
    \hline
    $m_{3/2}^{\rm NT} \ {\rm [GeV]}$& 509 & 179 & 76 \\
    \hline 
  \end{tabular}
  \caption{Typical mass parameters in each scenario. 
  In the last row we present the expected mass of the gravitino which explains 
  the whole of observed dark matter abundance through the production
  by the decay of other SUSY particles.
  }
  \label{tab:spectrum}
  \end{center}
\end{table}

\begin{figure}[tp]
\centerline{\epsfxsize=0.45\textwidth\epsfbox{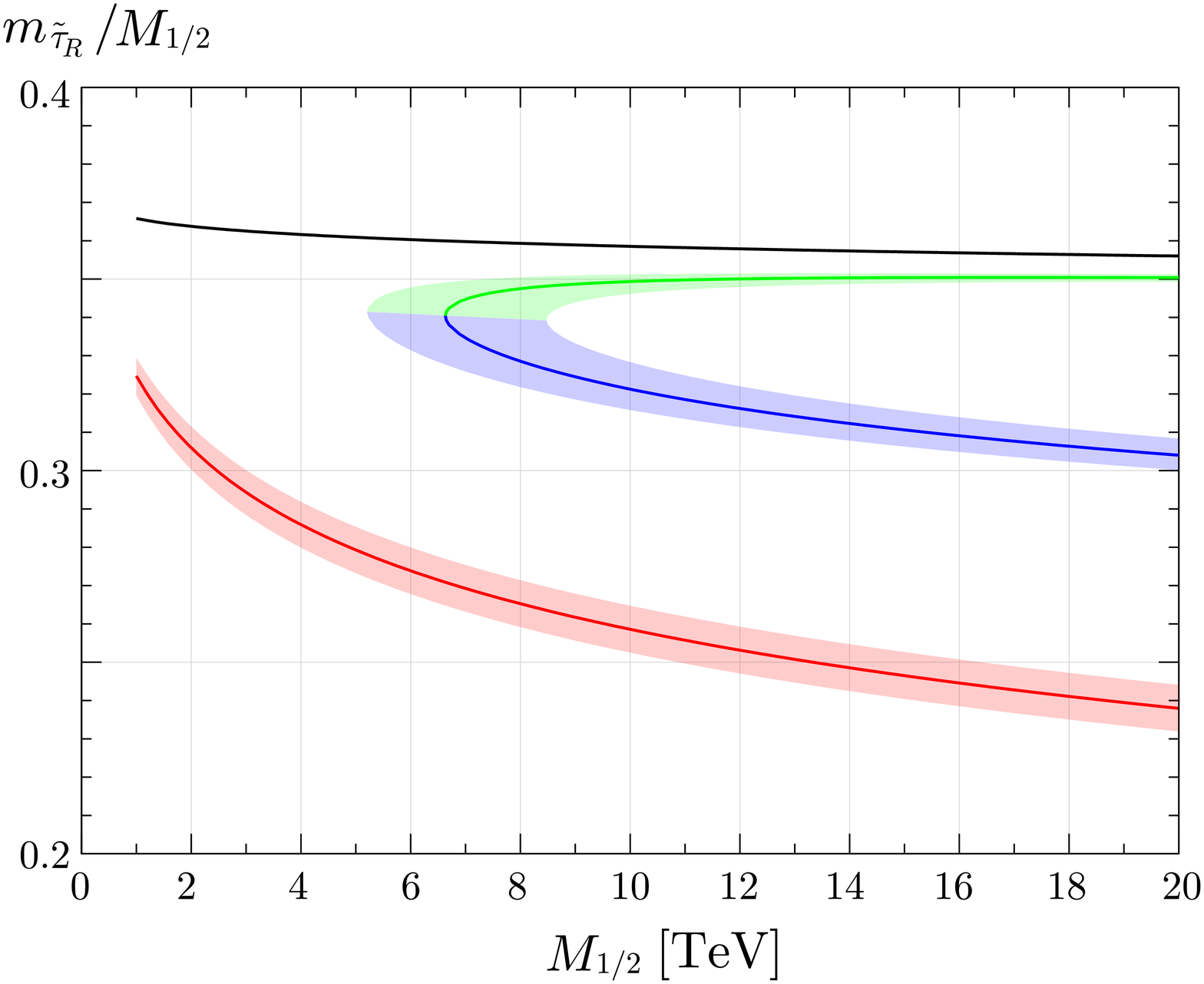}}
\caption{The right-handed stau mass normalized by $M_{1/2}$ as a function of $M_{1/2}$ 
 in the scenario \GI (red), \GII (blue) and \GIII (green). The black line represents the selectron mass 
 normalized by $M_{1/2}$. }
\label{fig:mstau}
\end{figure}

Since our model contains only one SUSY breaking scale which is much larger than EW scale,  
most of SUSY parameters are roughly proportional to $M_{1/2}$. 
Typical SUSY parameters for each solutions are presented in Table \ref{tab:spectrum}. 
The NLSP particle is the right-handed stau, which decays to the LSP gravitino dark matter. 
Stau mass is rather sensitive to $\tan\beta$, i.e., the tau Yukawa coupling 
since the flavor-independent contribution from gauge interactions 
is smaller than other sfermion mass terms.
In the case of the solution \GI~the right-handed stau mass is much smaller than selectron mass 
since the tau Yukawa coupling is effective. 
The mass difference between the stau and the selectron becomes smaller for the solution \GII~and 
they are almost degenerate in the case of the solution \GIII. 
It means that the right-handed stau gets heavier in the solution \GIII~compared to other 
solutions for a fixed $M_{1/2}$ as shown in Figure \ref{fig:mstau}. 
In the last row of Table \ref{tab:spectrum}, we present the gravitino mass which explains 
the observed dark matter abundance by the decay of other SUSY particles 
as will be explained in detail in the next section. 
This gravitino mass is roughly inversely proportional to $M_{1/2}$.


\section{Gravitino Dark Matter} \label{sec:gravitino}
As is explained in the introduction, the gravitino mass is assumed to be small to suppress 
the gravity-mediated contributions. Thus gravitino can be a good dark matter candidate in this model.

We assume that the NLSP stau is produced thermally in the history of the Universe and 
its thermal relic abundance reads as 
\begin{eqnarray}
\Omega_{{\tilde \tau}_R} h^2  \approx
 \frac{1.1 \times 10^9{\rm GeV}^{-1}}{\sqrt{g_{\star}} m_{\rm pl} \langle \sigma v \rangle x_f}
 \approx
  0.17 \left[ \frac{10} {\sqrt{g_{\star}}}\right]
  \left[ \frac{1/20}{x_f} \right]
  \left[ \frac{10^{-9}~{\rm GeV}^{-2}}{\langle \sigma v \rangle}\right]~,
  \label{eq:omegastau}
\end{eqnarray}
where $g_{\star}$ is the effective number of relativistic degrees of freedom at the freeze-out temperature $T_f$, 
$x_f = T_f/ m_{{\tilde \tau}_R}$ and $m_{\rm pl}=\sqrt{8\pi} M_{\rm pl}=1.22 \times 10^{19}~ {\rm GeV}$ . 
The right-handed staus mainly annihilate into the gauge boson pairs, 
${\tilde \tau} {\tilde \tau}^* \to \gamma \gamma$, $\gamma Z$, $ZZ$, $WW$, 
and also they annihilate into the tau lepton pairs through the Bino exchange, 
${\tilde \tau} {\tilde \tau} \to \tau \tau$ \cite{ttfinal}. 
Furthermore, 
since we find that the contribution through the left-right mixing, described below, cannot be neglected for large $\tan\beta$ case, we include the effects as denoted by $y_{\rm{eff}}$.
Then the thermally-averaged annihilation cross section of the right-handed staus is estimated as
\footnote{
Here we show only S-wave contributions and drop the terms suppressed 
by $\tan\beta$ and/or $m_{\rm SUSY}$. 
In the numerical calculation, we include remaining minor decay channels 
and solve the Boltzmann equations to include both the co-annihilation effects 
and P-wave contributions according to \cite{Griest:1990kh,Ellis:1999mm}. 
We have checked that Sommerfeld effects discussed in \cite{Berger:2008ti} give only negligible modifications in our set-up. 
}
\begin{eqnarray}
 \langle \sigma v \rangle \simeq 
 \frac{\pi \alpha^2}{m_{{\tilde \tau}_R}^2} \left[
   1+2 t_W^2 + \left\{ t_W^4 +\frac{(1 - y^2_{\rm eff})^2}{32 c_W^4}  \right\} + \frac{(1 - y^2_{\rm eff})^2}{16 c_W^4} 
 \right]
 + \frac{8\pi \alpha^2 M_{B}^2}{c_W^4 (m^2_{{\tilde \tau}_R}+M_B^2)^2}~.
 \label{eq:sigmav}
\end{eqnarray}
The each term in the parenthesis of the first term represent the contribution from 
$\gamma \gamma$, $Z\gamma$, $ZZ$ and $WW$ final states respectively. 
Here, $c_W = \cos \theta_W$, $t_W=\tan \theta_W$
and $y_{\rm eff} $ stands for the term proportional to the tau Yukawa coupling, 
\begin{eqnarray}
  y_{\rm eff} = 
    \frac{\mu}{ \sqrt{ m^2_{{\tilde \tau}_R} + m^2_{{\tilde \tau}_L}} } ~
    \frac{m_{\tau} \tan\beta}{t_W m_W (1+\varepsilon_\tau  \tan \beta )}~,
\end{eqnarray}
where $\varepsilon_\tau$ is the threshold corrections to the tau Yukawa coupling and 
$\varepsilon_\tau \simeq - 3\alpha_2/(16\pi) \mu M_2/m_{{\tilde \tau}_R}^2$ 
for degenerate SUSY masses.
This term is generated by the mixing effect between the left-handed and right-handed staus, 
and it becomes relevant for large $\tan\beta$, and it does not decouple in the limit of the large SUSY scale as one can see from the formula. Especially, it can remain significant in the limit of small left-right mixing, $\theta_{\tau}$, which is suppressed by the heavy SUSY mass, 
$ \theta_{\tau} \simeq {\cal O}( m_\tau \tan\beta / m_{\rm SUSY})$. 
On the contrary the enhancements of the annihilation into $hh$ \cite{Ratz:2008qh} and $tt$ \cite{Endo:2010ya} final states are irrelevant for us since they require both large $\tan\beta$ and large $\theta_{\tau}$. 

\begin{figure}[tp]
  \centerline{
  \epsfxsize=0.45\textwidth\epsfbox{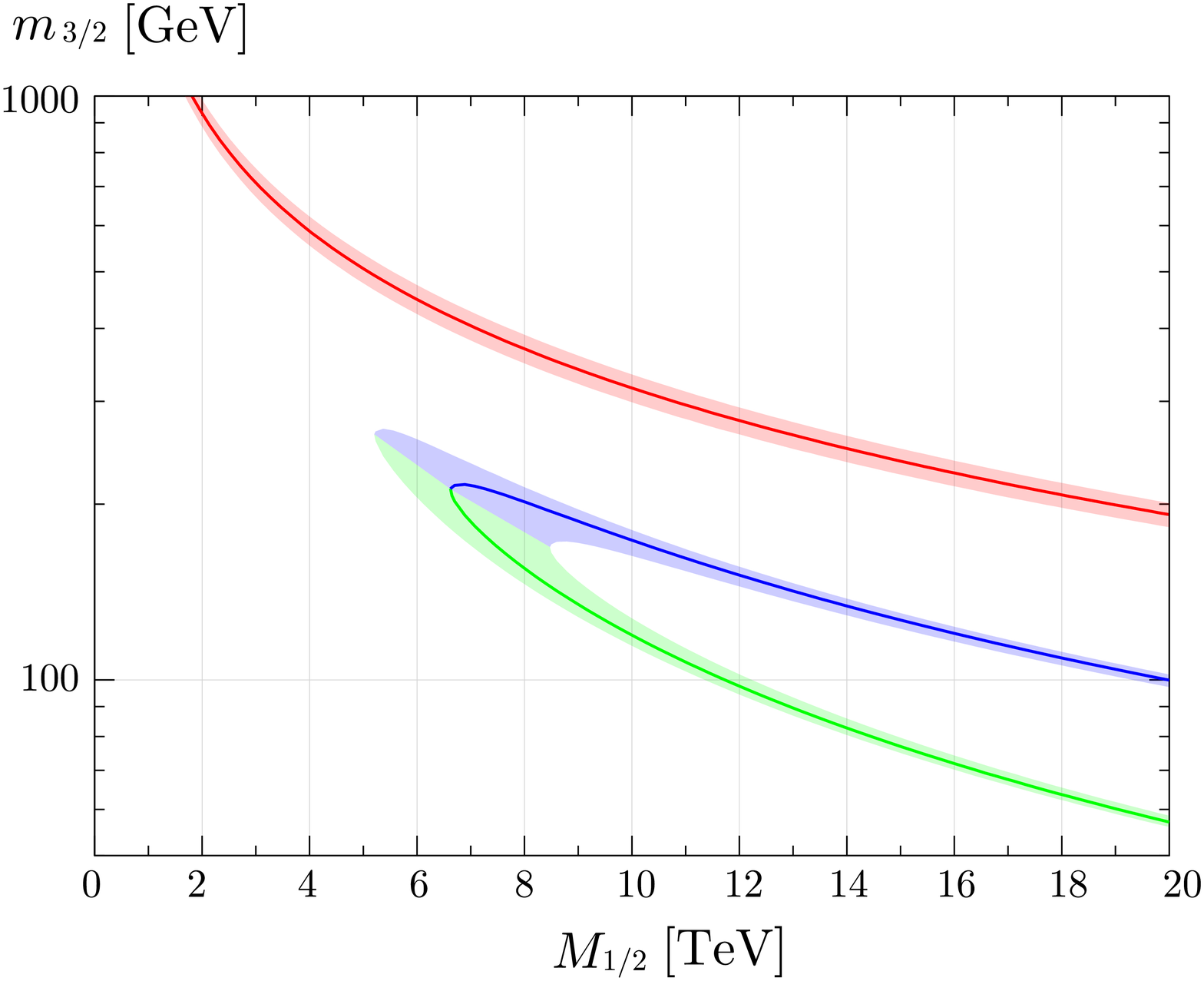}
  \epsfxsize=0.45\textwidth\epsfbox{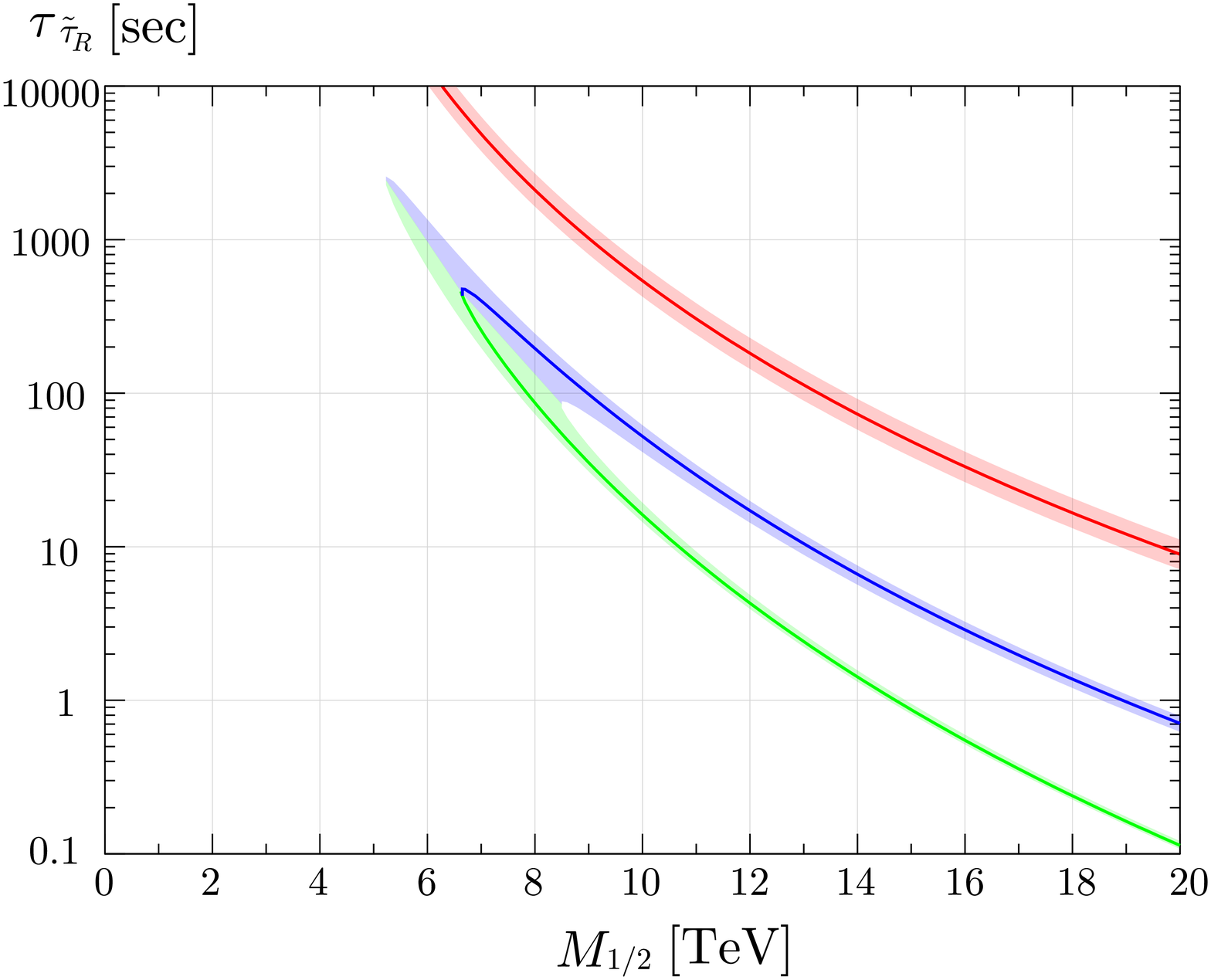}
  }
  \caption{ 
  The predicted mass of the gravitino LSP (left) and the life-time of the right-handed stau NLSP (right) 
  as functions of $M_{1/2}$ for the scenario \GI~(red), \GII~(blue) and \GIII~(green). 
  Here, the gravitino dark matters are assumed to be mainly produced by the decay of other SUSY particles
  and become the dominant component of the cold dark matter in our Universe. 
  }
  \label{fig:m32}
\end{figure}

The gravitinos are generated non-thermally through the decay of the staus.
The relic density of the gravitino dark matter is given as
\begin{eqnarray}
 \Omega^{ \rm NT}_{\tilde G} h^2 =  \frac{m_{3/2}}{m_{{\tilde \tau}_R}} \Omega_{{\tilde \tau}_R} h^2~.
\end{eqnarray}
The minimal scenario is to consider that the observed cold dark matter density $\Omega_c h^2 = 0.12$ \cite{1502.01589} 
is explained by the gravitino dark matter produced in this way. 
Thus, we can predict the gravitino mass as  
$m^{\rm NT}_{3/2} = m_{{\tilde \tau}_R} \Omega_c/ \Omega_{{\tilde \tau}_R}$, 
which is roughly inversely proportional to the right-handed stau mass 
from Eqs. (\ref{eq:omegastau}) and (\ref{eq:sigmav}). 
In Figure \ref{fig:m32} (left), we show the predicted masses of gravitino dark matter for each scenarios. 
We find the gravitino masses sit in the right parameter range anticipated by a naive estimation, 
$m_{3/2}/M_{1/2} \approx {\cal O}(M_{\rm G}/M_{\rm pl}) \simeq {\cal O}(0.01)$, 
and it gives the strong implication on the concrete ultraviolet model construction. 
In the scenario \GIII, the gravitino can become relatively light, 
below $100~{\rm GeV}$ for $M_{1/2} > 12~{\rm TeV}$, 
since the right-handed stau mass tends to be heavy due to the small $\tan\beta$. 
In the following we will see that light gravitinos are favored by the constraints from the BBN 
and also from the EDM experiment. 
Then, in the scenarios \GI~and \GII, the whole of the dark matters cannot be identified 
as non-thermally produced gravitinos
and we must consider other production mechanisms and/or other dark matter candidates.
In a later section, we will consider the case where gravitinos are also thermally 
produced in the early stage of the Universe
and discuss the implications on the reheating temperature after the inflation. 

Next we consider the BBN constraints on this model. 
Since the life-time of the stau NLSP is relatively long in our model, its late-time decay 
may destroy the successful predictions of the standard BBN scenario \cite{0404198,0408426}.
The stau life-time is estimated as 
\begin{eqnarray}
 \tau_{{\tilde \tau}_R} \simeq \frac{48\pi m_{3/2}^2 M_{\rm pl}^2}{m^5_{{\tilde \tau}_R}}
 \simeq 10\ {\rm sec} \left(\frac{m_{3/2}}{100\ {\rm GeV}}\right)^2 \left( \frac{m_{{\tilde \tau}_R}}{3\ {\rm TeV}} \right)^{-5}~.
\end{eqnarray}
For $\tau_{{\tilde \tau}_R} < 1\ {\rm sec}$, the NLSP stau can decay before the BBN starts.
For $1\ {\rm sec} < \tau_{{\tilde \tau}_R} < 100\ {\rm sec}$, 
the hadronic particles produced by the NLSP decay are potentially dangerous 
since they interconvert the protons and neutrons and may change the helium abundance. 
Especially in the stau NLSP case, since the staus mainly decay to the tau lepton and gravitino pairs,  
${\tilde \tau}_R \rightarrow {\tilde G} \tau$, 
sizable amounts of pions are produced through the hadronic decay of tau leptons. 
However the produced pions are less effective to interconvert the protons and neutrons than $p {\bar p}$
and $n {\bar n}$ pairs. 
From the figures of \cite{0804.3745}, we find that the stau abundance considered here is not problematic. 
Although these analyses are performed for lower stau masses, produced taus are stopped immediately 
with losing their energy electromagnetically and therefore the constraint is less sensitive to the stau mass itself 
in this time range. 
On the other hand, the case with $ \tau_{ {\tilde \tau}_R} > 100\ {\rm sec}$ is excluded by the overproduction of D as one can see from Ref. \cite{0804.3745}.
Then, in the following we assume that thermally produced NLSP staus are allowed as long as 
$ \tau_{ {\tilde \tau}_R} < 100\ {\rm sec}$, 
and clearly more detailed analyses of BBN constraints on heavy NLSP masses are desirable. 

We show the stau life-time as a function of $M_{1/2}$ in Figure \ref{fig:m32} (right).
Comparing with the predictions of the Higgs mass (Figure \ref{fig:mHiggs}), we can conclude that the parameter region explaining the correct Higgs mass
is in tension with the BBN constraint in the scenario \GI. 
In the scenario \GII~we need $M_{1/2} \gsim 9~{\rm TeV}$ and the top quark mass is preferably a little bit smaller 
than the currently measured central value, and a further wider parameter region survives in the scenario \GIII. 

\section{Electric Dipole Moments}  \label{sec:edms}
Although we assume the favor- and CP-safe condition Eq.(\ref{eq:safe}) at the tree-level, 
the additive gravity mediation effects of the order of the gravitino mass are not negligible in our model if the abundance of the 
gravitino dark matter is explained by the decay of NLSP stau particles. 
Such gravity-mediated contributions become new sources 
of the flavor- and CP-violation. 

Especially in our model, since the $B$ parameter is as small as the gravitino mass, 
the gravity-mediated contribution to the $B$ parameter can lead to 
a large CP-violating phase of $\mu$ term in the basis where Higgs VEVs and gaugino masses are real. 
Then we expect visible effects on the EDM measurements
\footnote{
The gravity-mediated contributions to the A term can be another sources of the CP violation. 
Here we assume that their sizes are somehow controlled by the corresponding Yukawa couplings 
in order to avoid the color and charge breaking minimum.
With this assumption, the dominant contribution to the EDM comes from the phase of $\mu$ term 
thanks to the enhancement of $\tan\beta$. 
}.
In the following we assume that the phase of $\mu$ term is shifted by the gravity-mediated contribution as
\begin{eqnarray}
  \phi_{\mu} = \phi_{\mu}^{(0)} + {\rm max} \left\{  \frac{m_{3/2}} {|B|}, \frac{\pi}{2} \right\},~
  \label{eq:CP}
\end{eqnarray}
where the $\phi_{\mu}^{(0)}$ is the phase without the gravity mediated contributions;
$\phi_{\mu}^{(0)}=0$ for the scenario {\GI} and $\phi_{\mu}^{(0)}=\pi$ for the scenario {\GII} and {\GIII}. 

The leading SUSY contributions to the electron EDM come from the loop diagrams with chargino-sneutrino 
and neutralino-selectron exchange. They are well approximated as 
\begin{eqnarray}
  \frac{d_e}{e} =  \frac{\alpha_2}{4\pi} \frac{m_e t_\beta}{m_{{\tilde e}_L}^4} 
    \frac{ |M_2 \mu| \sin \phi_\mu }{1+\varepsilon_e^2 t_\beta^2 + 2 \varepsilon_e t_\beta \cos\phi_\mu}
   \left[ 
   F^{(e)}_2 (x_{2L}, x_{\mu L}) + \frac{\alpha_Y |M_1|}{\alpha_2 |M_2|} F^{(e)}_1 (x_{1L}, x_{\mu L}, x_{1R}, x_{\mu R}) 
   \right] ,
   \label{eq:de}
\end{eqnarray}
with $t_\beta=\tan\beta$, $x_{1L/1R} = |M_1|^2/ m^2_{{\tilde e}_{L/R}}$, $x_{2L/2R} = |M_2|^2/ m^2_{{\tilde e}_{L/R}}$ 
and $x_{\mu L/\mu R} = |\mu|^2/ m^2_{{\tilde e}_{L/R}}$.
The loop functions $F^{(e)}_2$ and $F^{(e)}_1$ are defined in Appendix \ref{app:functions}. 
The coefficient $\varepsilon_e$ stands for the $\tan\beta$-enhanced threshold corrections to the electron mass and it reads
\begin{eqnarray}
  \varepsilon_e &=&  -\frac{3 \alpha_2}{16\pi} \frac{|\mu M_2|}{m^2_{{\tilde e}_L}} 
 \left[  I_2^{(e)}(x_{2 L},x_{\mu L})  + \frac{\alpha_Y |M_1|}{\alpha_2 |M_2|} 
 I_1^{(e)}(x_{1L},x_{\mu L},x_{1R},x_{\mu R}) 
\right]~,
\end{eqnarray}
where the loop functions $I_2^{(e)}$ and $I_1^{(e)}$ are listed in Appendix \ref{app:functions}.
In the limit of the common SUSY breaking masses, we find $F_2^{(e)}(1,1) =-5/24$, $F_1^{(e)}(1,1,1,1) =-1/24$  and
$I_2^{(e)}(1,1) = 1$, $I_1^{(e)}(1,1,1,1)=-1/3$. 

\begin{figure}[tp]
  \centerline{\epsfxsize=0.45\textwidth\epsfbox{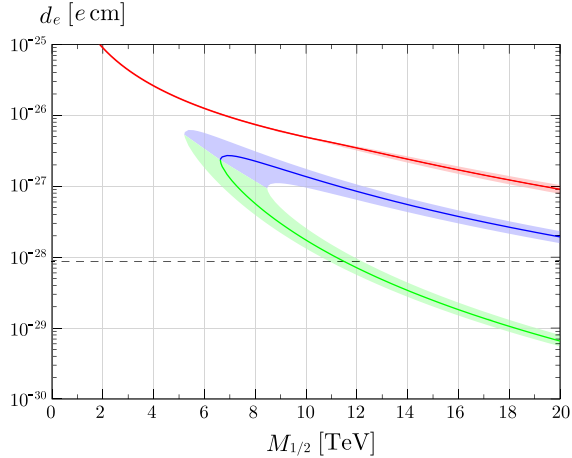}}
  \caption{
  The predicted electron EDMs as a function of $M_{1/2}$ in the scenario \GI~(red), \GII~(blue) and \GIII~(green). 
  The dotted line shows the current experimental upper bound \cite{Baron:2013eja}.
  }
  \label{fig:de}
\end{figure}

In Figure \ref{fig:de}, we show the expected values of the electron EDM, assuming the CP phase as Eq. (\ref{eq:CP}). 
We find that the whole of the parameter regions that explain the observed Higgs mass are disfavored 
by the EDM measurement in the scenario \GI~and \GII. 
On the other hand, the constraint becomes milder for the scenario \GIII~thanks to 
small $\tan\beta$ and small $\sin\phi_\mu \simeq m_{3/2}/|B|$. 
Although these predictions of the electron EDM contain {\cal O}(1) uncertainty because the exact size 
of gravity-mediated contributions are unknown, it is plausible that the Higgs mass is explained 
in the scenario \GIII~with $10~{\rm TeV} \lsim  M_{1/2} \lsim 20~{\rm TeV}$ 
and the electron EDM is expected to be detected in the near future experiments.

Since the SUSY spectrum is controled by a common parameter, $M_{1/2}$,
and thus one can obtain rigid predictions for the ratios between electron
and quark (C)EDMs irrespective of the actual size of the CP-violating
phase.  In this model, dominant contributions to the hadronic EDMs comes
from down quark (C)EDMs, and it reads
\begin{eqnarray}
  \frac{d_d}{e} &=&  \frac{\alpha_3}{4\pi} \frac{m_d t_\beta}{m_{{\tilde d}_L}^4 } 
   \frac{ |M_3 \mu| \sin \phi_\mu }{1+\varepsilon_d^2 t_\beta^2 + 2 \varepsilon_d t_\beta \cos\phi_\mu}
   \bigg[ 
   F^{(d)}_3 (y_{3L}, y_{3R})
    + \frac{\alpha_2 |M_2|}{\alpha_3 |M_3|} F^{(d)}_2 (y_{1L}, y_{\mu L}) 
       \nonumber\\
    &&
    \quad + \ \frac{\alpha_Y |M_1|}{\alpha_3 |M_3|} F^{(d)}_1 (y_{1L}, y_{\mu L}, y_{1R}, y_{\mu R}) 
   \bigg] ,
   \label{eq:dd}
   \\
  d^c_d &=&  \frac{\alpha_3}{4\pi} \frac{m_d t_\beta}{m_{{\tilde d}_L}^4 } 
  \frac{ |M_3 \mu| \sin \phi_\mu }{1+\varepsilon_d^2 t_\beta^2 + 2 \varepsilon_d t_\beta \cos\phi_\mu}
  \bigg[ 
   G^{(d)}_3 (y_{3L}, y_{3R})
    + \frac{\alpha_2 |M_2|}{\alpha_3 |M_3|} G^{(d)}_2 (y_{1L}, y_{\mu L}) 
    \nonumber\\
    &&
    \quad +\  \frac{\alpha_Y |M_1|}{\alpha_3 |M_3|} G^{(d)}_1 (y_{1L}, y_{\mu L}, y_{1R}, y_{\mu R}) 
   \bigg] ,
   \label{eq:dcd}
\end{eqnarray}
where the $\tan\beta$-enhanced threshold corrections to down quark mass is parametrized by 
\begin{eqnarray}
  \varepsilon_d  \!\!\! &=& \!\!\!   
  \frac{\alpha_3}{3\pi} \frac{|\mu M_3|}{m^2_{{\tilde d}_L}} 
 \bigg[  
 I_3^{(d)}(y_{3L},y_{3R})  
 + \frac{\alpha_2|M_2|}{\alpha_3 |M_3|}  I_2^{(d)}(y_{2L},y_{\mu L})  
 + \frac{\alpha_Y|M_1|}{\alpha_3 |M_3|}  I_1^{(d)}(y_{1L},y_{\mu L},u_{1R},y_{\mu R}) 
\bigg]~.
\end{eqnarray}
Here, $y_{1L/1R} = |M_1|^2/ m^2_{{\tilde d}_{L/R}}$, $y_{2L/2R} = |M_2|^2/ m^2_{{\tilde d}_{L/R}}$ 
and $y_{\mu L/\mu R} = |\mu|^2/ m^2_{{\tilde d}_{L/R}}$.
The loop functions $F^{(d)}_i$, $G^{(d)}_i$ and $I^{(d)}_i$ $(i=1,2,3)$ are defined in Appendix \ref{app:functions}. 
For the common SUSY breaking masses, the functions become 
$F^{(d)}_{\{1,2,3\}}= \{11/648,-7/24,-2/27 \}$, 
$G^{(d)}_{\{1,2,3\}}=\{-11/216,-1/8,-5/18\}$ and
$I^{(d)}_{\{1,2,3\}}=\{-11/48,-9/16, 1\}$. 
Similarly the CP phase also generates the strange quark (C)EDMs, which is estimated as
$d_s/d_d \simeq d^c_s/d^c_d \simeq m_s/m_d \simeq 18$, but  their contributions to 
the hadronic EDMs are still uncertain \cite{Hisano:2012sc}. 
These quark (C)EDMs are evaluated at the SUSY scale and their values at the hadronic scale, 
$\mu_H = 1~{\rm GeV}$,  are obtained by the renormalization evolution \cite{Degrassi:2005zd}. 

\begin{figure}[tp]
  \centerline{
  \epsfxsize=0.45\textwidth\epsfbox{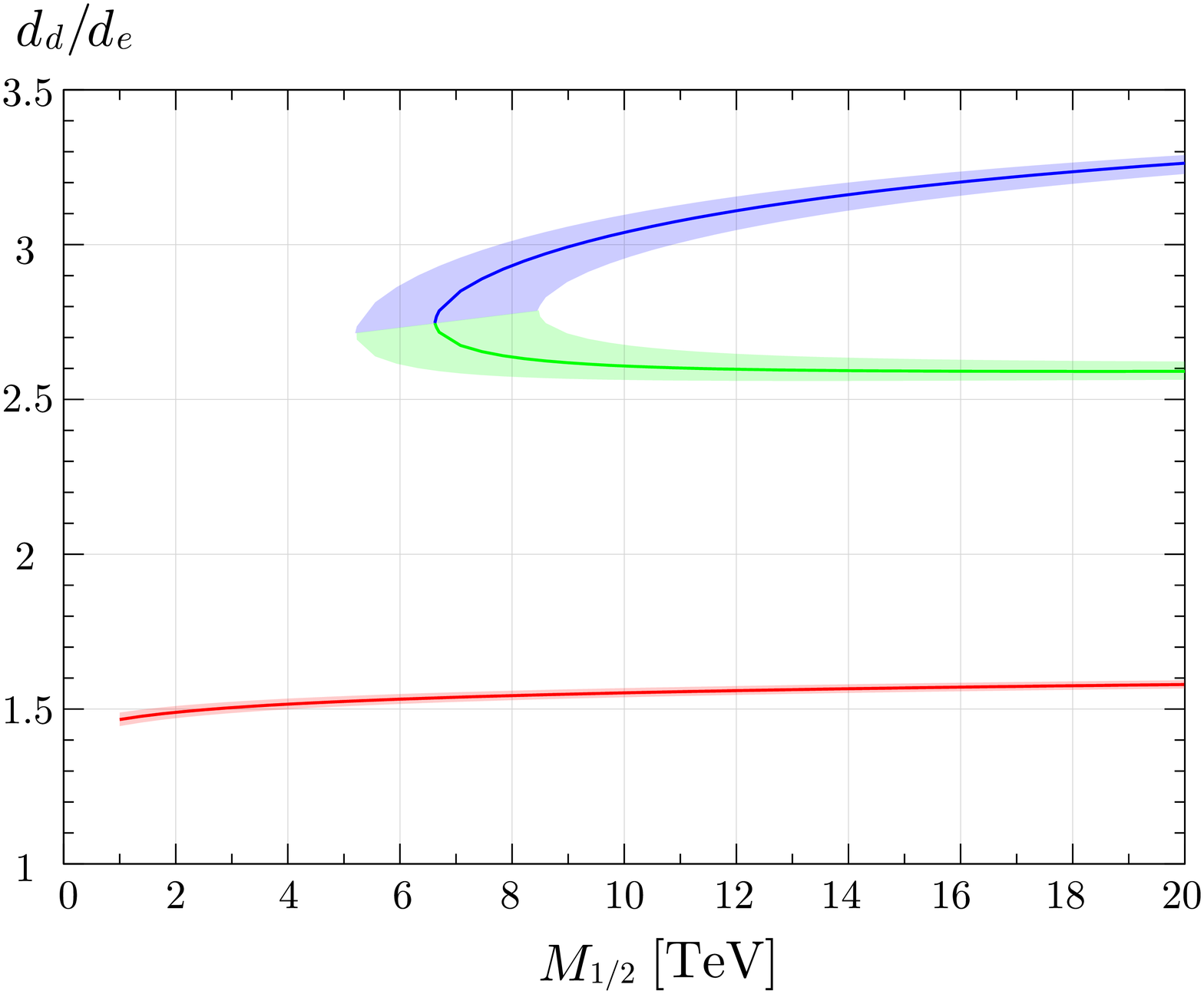}
  \epsfxsize=0.45\textwidth\epsfbox{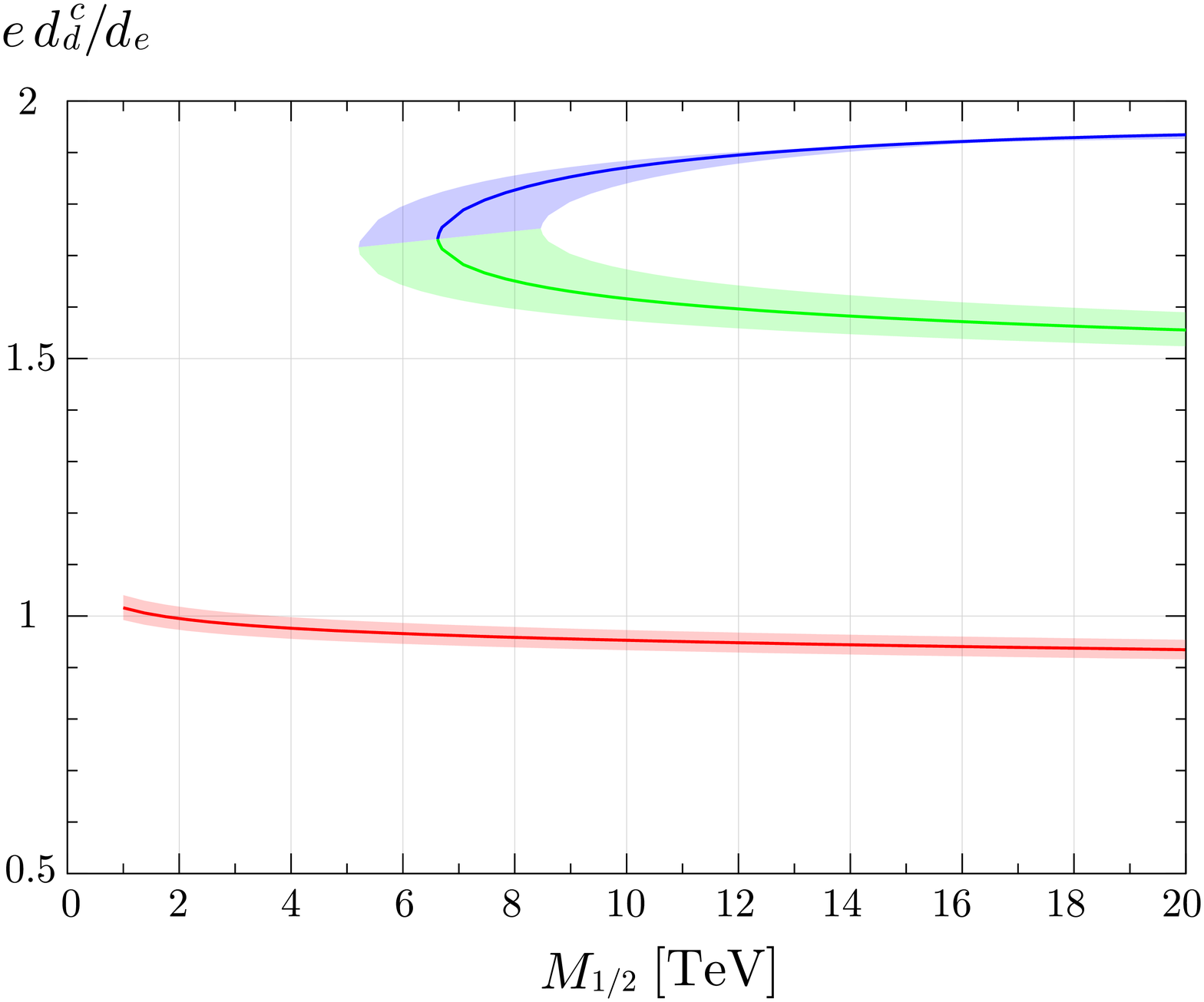}
  }
  \caption{Predicted ratios of down-quark EDM (left) and CEDM (right) to the electron EDM as a function of $M_{1/2}$ 
  in the scenario \GI~(red), \GII~(blue) and \GIII~(green).
  }
  \label{fig:dd}
\end{figure}

From Eqs.~(\ref{eq:de}), (\ref{eq:dd}) and (\ref{eq:dcd}), the electron EDM and down quark (C)EDMs are 
almost proportional to the imaginary part of the phase, $\sin \phi_\mu$, and their ratios have rather weak dependence
on the size of the phase. 
In the scenarios \GI~and \GII~the naive expectation of the phase can be ${\cal O}(1)$. 
However this uncertainty becomes almost negligible once the size of the phase is restricted 
to satisfy the current experimental bound. 
In Figure \ref{fig:dd}, we show the predicted ratios of the down quark (C)EDMs to the electron EDM, 
with setting the CP phase to be small enough to satisfy the experimental bound. 
For $\cos\phi_\mu < 0$, the down quark (C)EDMs are enhanced by the threshold corrections 
to the down quark mass term while the electron EDM is decreased. 
Therefore the down quark (C)EDMs can become relatively large in the scenario \GII~and \GIII. 

The experimental sensitivities of the electron EDM \cite{Sakemi:2011zz,Kara:2012ay,Kawall:2011zz} and the nucleon 
EDMs \cite{Piegsa:2013vda,Matsuta:2013usa,Serebrov:2015idv,Semertzidis:2011qv,Lehrach:2012eg} are expected 
to be improved by several orders of magnitudes in the next-generation experiments, 
and the measurements of these EDMs is essential to confirm or reject our model.
The nucleon EDMs are induced by the quark (C)EDMs and 
their contributions are estimated by using the QCD sum rules as follows \cite{Hisano:2012sc,Hisano:2015rna}
\footnote{
This formulae are obtained with the assumption that the Peccei-Quinn symmetry 
works to suppress the contribution from the QCD $\theta$ term.
}, 
\begin{eqnarray}
 d_n \!\!\!&=&\!\!\! -0.20 d_u + 0.78 d_d + e ( 0.29 d^c_u + 0.59 d^c_d)~,
 \label{eq:dn}
 \\
 d_p \!\!\!&=&\!\!\! 0.78 d_u - 0.20 d_d + e ( -1.2 d^c_u - 0.15 d^c_d)~. 
  \label{eq:dp}
\end{eqnarray}
Note, while the quark EDM contributions to the neutron EDM are well consistent with the recent result
obtained by the lattice simulation \cite{Bhattacharya:2015esa}, the quark CEDM contributions contain large theoretical uncertainties 
and they should be fixed ultimately by the lattice QCD calculation. 
With the expressions of Eqs.~(\ref{eq:dn}) and (\ref{eq:dp}), 
the sizes of nucleon EDMs are predicted as 
$d_n/d_e \simeq1.8,~3.5,~3.0$ and 
$d_p/d_e \simeq -0.5,~-0.9,~-0.8$ for $M_{1/2}=15\ {\rm{TeV}}$\footnote{Since $M_{1/2}$ dependence of $d_n/d_e$ and $d_p/d_e$ is not so significant, we have shown values at the sample point. } in the scenario \GI, \GII~and \GIII~respectively.


%
%
\section{Thermally Produced Gravitinos} \label{sec:thermal}
As we have seen in earlier subsections, if most of the cold dark matter in our Universe consist 
of the gravitinos produced by the decay of other SUSY particles, the scenarios \GI~and \GII~are 
disfavored by the BBN constraint and/or the EDM measurement. These constraints can be alleviated 
if the gravitino is much lighter than those expected by the calculation of the non-thermal production.  

In the actual ultraviolet physics, it is non-trivial whether light gravitinos can be realized with satisfying 
the boundary conditions Eqs.~(\ref{eq:safe}) and (\ref{eq:safe2}) at the GUT scale. 
For example, the models using the gaugino-mediation are primary candidates of our UV complete model, 
but the gravitino mass has lower bounds depending on the number of extra dimensions \cite{Buchmuller:2005rt}. 

Here we leave aside such difficulty of the model-building, and discuss phenomenological implications 
of the model with much lighter gravitinos.
In this case non-thermally produced gravitinos cannot be the main constituent of the cold dark matter. 
The gravitinos can be also produced thermally by scattering processes with particles in the thermal bath.
The resultant abundance is approximately proportional to the reheating temperature after the inflation, $T_R$, 
and it reads \cite{0804.3745}
\begin{eqnarray}
 \Omega_{\tilde G}^{\rm th} h^2 \simeq 0.41
 \left( \frac{m_{3/2}}{100~{\rm GeV}} \right)^{-1}
 \left( \frac{M_{1/2}}{10~{\rm TeV}} \right)^2
 \left( \frac{T_{R}}{10^7~{\rm GeV}} \right)~. 
\end{eqnarray}
Requiring the total gravitino abundance is smaller than the observed dark matter abundance, 
$\Omega^{\rm NT}_{\tilde G}+\Omega^{\rm th}_{\tilde G} < \Omega_c$, 
we obtain the upper bounds on the reheating temperature. 

\begin{figure}[tp] 
  \centerline{
  \epsfxsize=0.45\textwidth\epsfbox{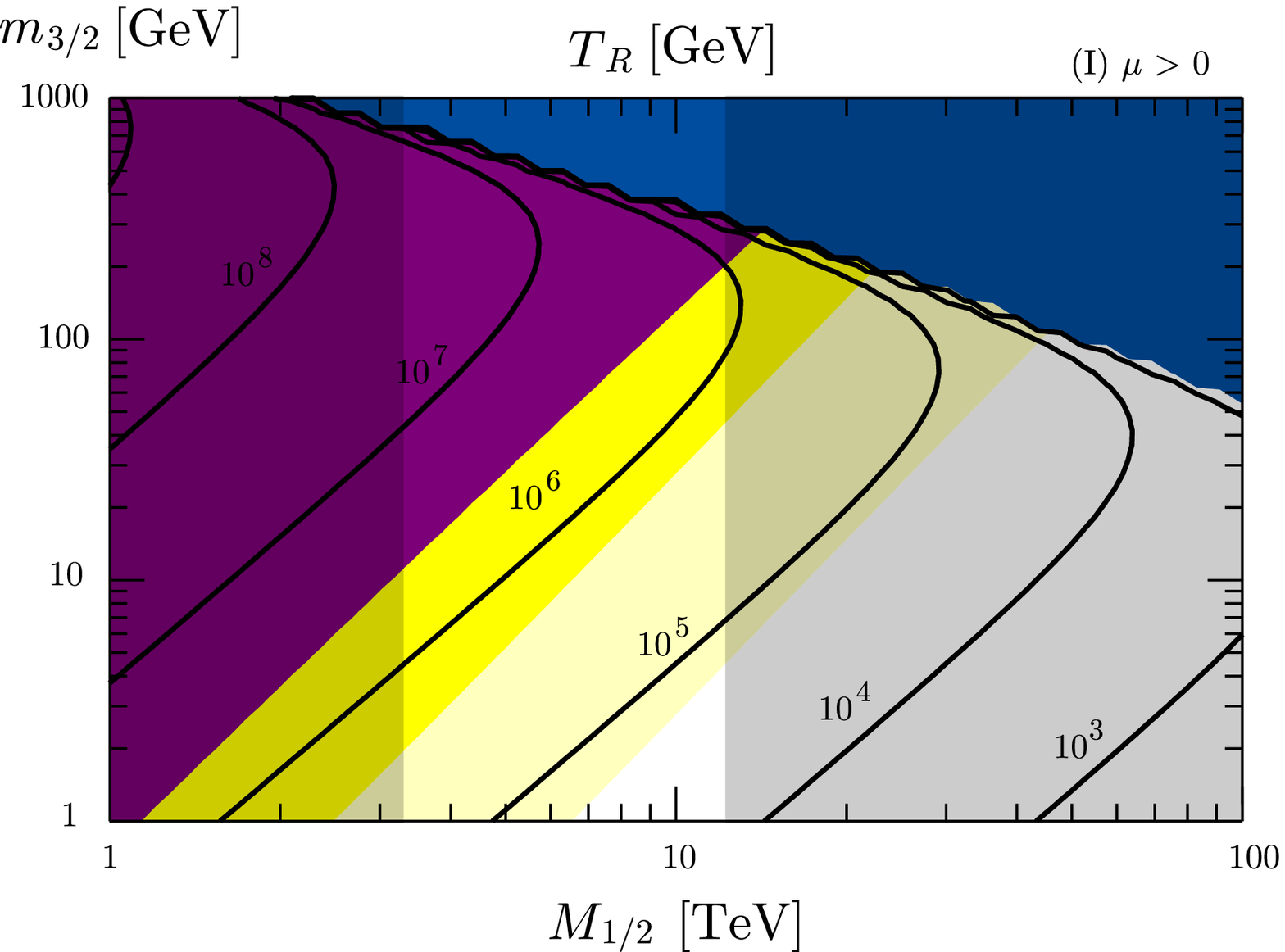}
  \epsfxsize=0.45\textwidth\epsfbox{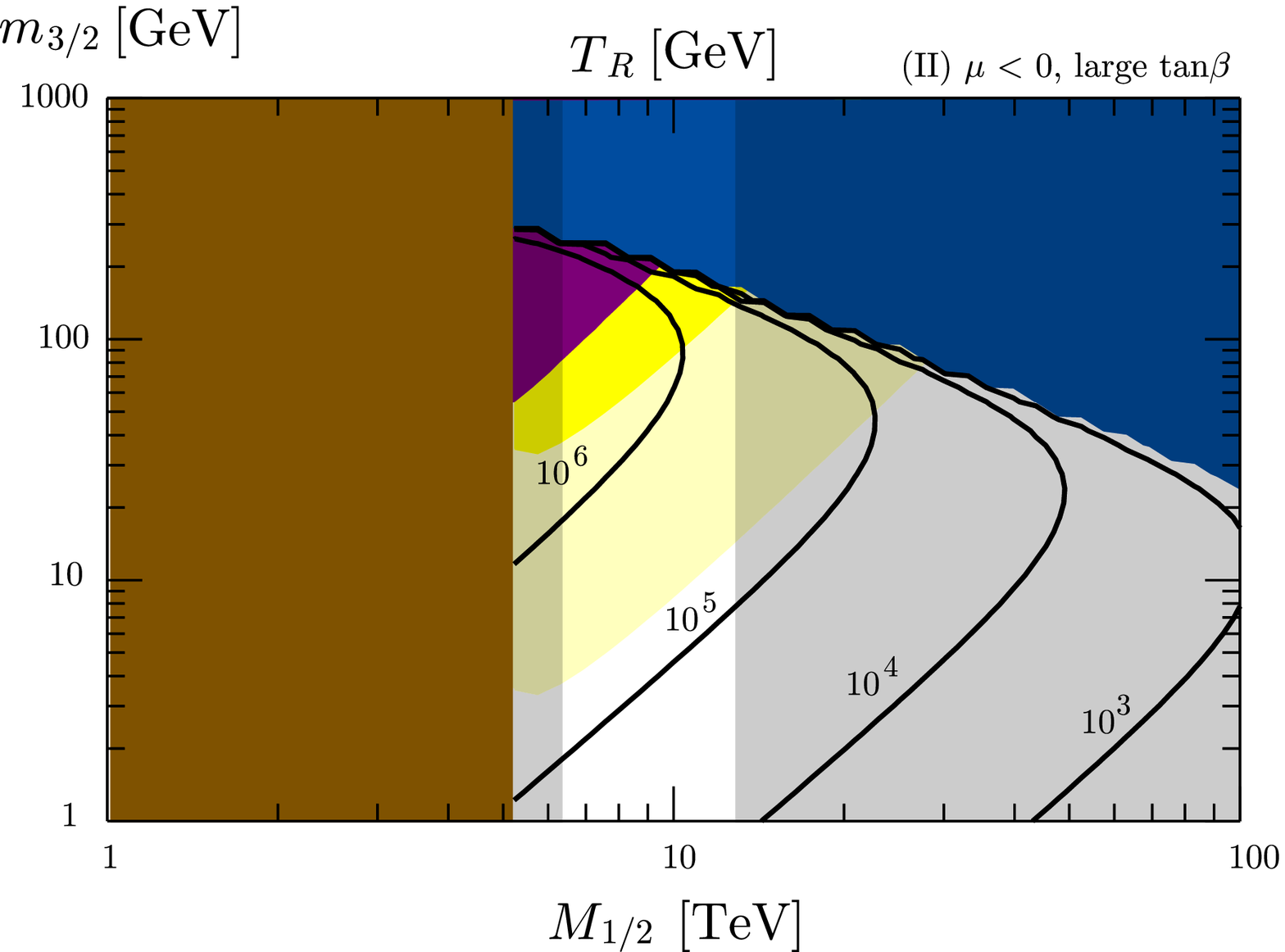}
  }
  \vspace{0.5cm}
  \centerline{
  \epsfxsize=0.45\textwidth\epsfbox{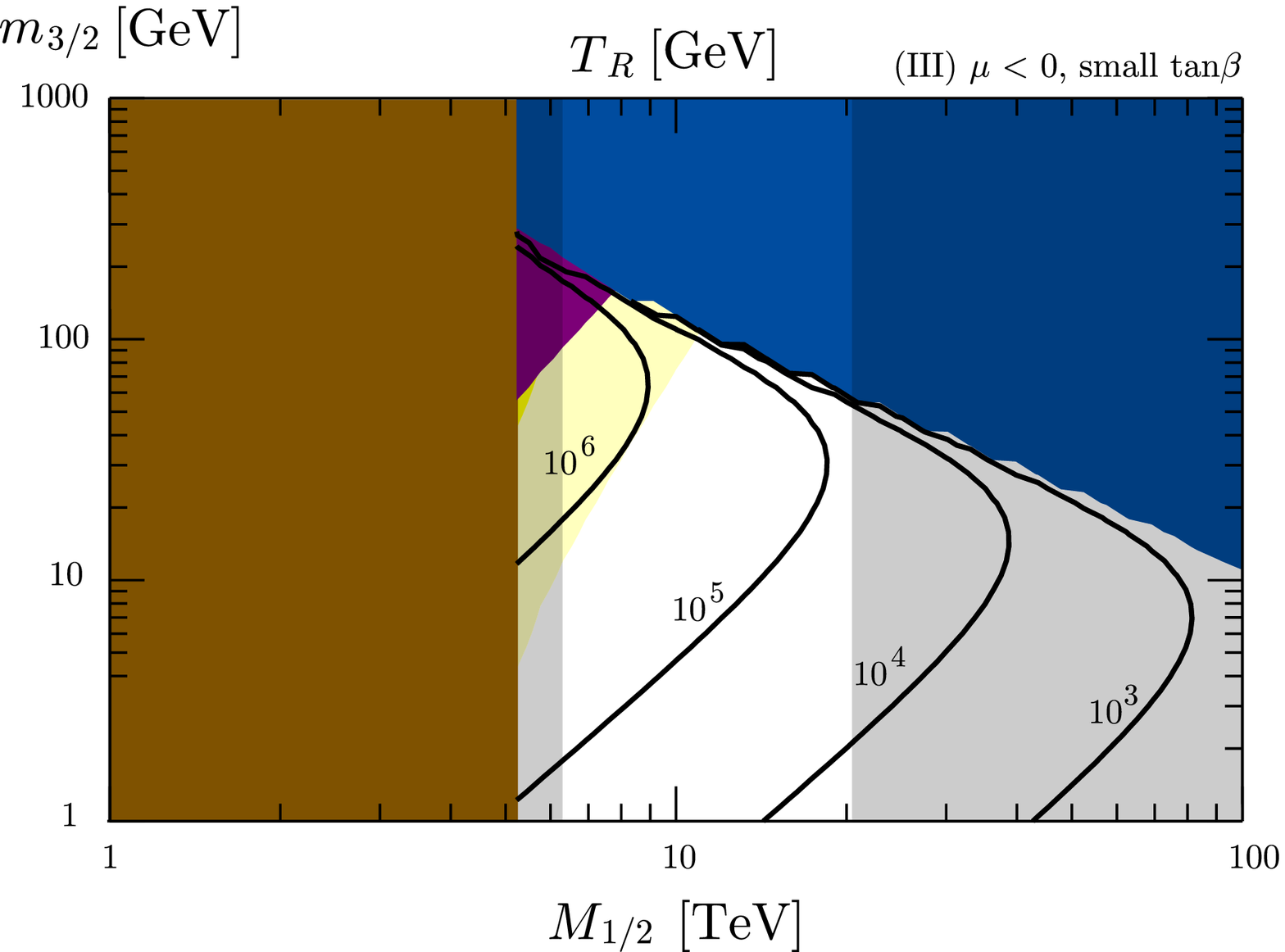}
  }
  \caption{upper bounds on the reheating temperature, $T_R$, as functions of $M_{1/2}$ 
   and $m_{3/2}$ in the scenario \GI~(top left), \GII~(top right) and \GIII~(bottom). 
   Colored regions are excluded by the following conditions, 
   {\bf Gray:} explanation of the Higgs mass,
   {\bf Blue:} overabundance of the gravitino dark matter relic density,
    {\bf Brown:} no EWSB minimum,
   {\bf Purple:} BBN constraints,
   {\bf Yellow:} no observation of the electron EDM, assuming $\phi_B= m_{3/2}/B$ (light yellow)
   and $\phi_B=0.1~m_{3/2}/B$ (dark yellow).  }
  \label{fig:TR}
\end{figure}

In Figure \ref{fig:TR} we show the upper bounds on the reheating temperatures in the 
$(M_{1/2}, m_{3/2})$ plane for the scenario \GI, \GII~and \GIII. 
Gray shaded regions cannot explain the Higgs mass even if the top mass is 
chosen in the range of $M_t = 173.3 \pm 2~{\rm GeV}$. 
The gravitino abundance exceeds the observed dark matter abundance in the blue regions.  
Purple regions are excluded by the BBN constraints, which imply $\tau_{\tilde \tau} > 100\ {\rm sec}$, 
and the correct EWSB minimum cannot be obtained in the brown regions. 
Yellow regions are disfavored by the electron EDM constraint, assuming $\phi_B= m_{3/2}/B$ (light yellow)
and $\phi_B=0.1~m_{3/2}/B$ (dark yellow).
Apart from the calculation of the Higgs mass we choose $M_t = 171.3~{\rm GeV}$ in this plot, 
and its precise value is less sensitive to the predictions of the reheating temperature and other constraints.
 
We find that the gravitino mass should be less than about 10 GeV in the scenario \GI~and \GII~
to suppress the SUSY contributions to the electron EDM.
With this small gravitino mass the reheating temperature becomes relatively low, 
less than about $10^5$ GeV.
In the scenario \GIII~the constraint from the EDM  becomes weaker, and the upper bound on the reheating 
temperature is raised up to $10^6$ GeV.
 
\section{Summary and discussion} \label{sec:summary}
In this paper we have discussed a model where the SUSY breaking effects are mediated only to the gaugino 
mass terms at the leading order as a simple solution for the SUSY flavor and CP problems.
The gravitino mass should be smaller than other SUSY breaking parameters to suppress the 
gravity-mediated contributions, which are the main sources of flavor and CP violation in our model.
Thus, in the presence of the R-parity conservation, the gravitino becomes a good candidate 
of the cold dark matter in our Universe.

We carefully examine the RG running of the $B$ parameter and find that the parameter 
region with small $\tan\beta$ appears as the SUSY scale increases. 
Thanks to the smallness of $\tan\beta$, 
the SUSY breaking scale to explain the Higgs boson mass is pushed up, 
and especially the NLSP right-handed stau mass is increased.
Then, since the life-time of the NLSP staus is shorten in this parameter region,  
the whole of the observed dark matter abundance can be explained 
by the gravitinos produced by the decay of other SUSY particles 
without destroying the successful predictions of the standard BBN
scenario.

Since the $B$ parameter is relatively small compared to other SUSY breaking parameter, 
it is affected by the small gravity-mediated contributions. 
Thus we expect non-negligible SUSY contributions to the electron and nucleon EDMs.
We find that the naively expected sizes of the electron EDM are on the edge of the current experimental limit, 
and it should be checked in the near future experiments. 
Especially the ratios of the nucleon EDMs to the electron EDM become the key ingredients 
for the discrimination of our model with others. 

Our model is very predictive and the whole of the SUSY mass spectrum is fixed
if the top Yukawa coupling is measured precisely and the theoretical errors in the Higgs mass 
calculation are reduced. 
Then it is important to measure the mass of the lightest MSSM SUSY particle, right-handed stau, 
at the future collider experiments 
%
such as in the future $100~{\rm TeV}$ hadron colliders \cite{Cohen:2013xda,Feng:2015wqa}. 
We expect that our model would be tested by the combination of the future collider and EDM experiments.

Finally we comment on the UV theory of the model. 
In this analysis the $\mu$ term is assumed to be a fundamental parameter which is comparable to the SUSY scale 
and tuned to generate the correct electroweak scale. 
And also, the gravitino mass is treated as a free parameter. 
To understand the whole picture of our model, it is desirable to construct the UV model 
which explains the origin of $\mu$ term and predicts the preferable gravitino mass.

\vspace{1em}
\noindent {\it Acknowledgements}: 

This work is supported by JSPS KAKENHI Grant-in-Aid for Scientific Research (B) (No. 15H03669 [RK])
, MEXT Grant-in-Aid for Scientific Research on Innovative Areas (No. 25105011 [RK]).

\appendix
\section{Loop Functions} \label{app:functions}
We list the loop functions relevant to the calculation of EDMs, 
\begin{eqnarray}
   F^{(e)}_1 (x_{1L}, x_{\mu L}, x_{1R}, x_{\mu R}) \!\!\!&=&\!\!\!
   \frac{1}{4} D_0(x_{1L}, x_{\mu L})
   - \frac{x_{1R}^2}{2 \, x_{1L}^2} D_0(x_{1R}, x_{\mu R}) 
   - \frac{x_{1R}}{2\,  x_{1L}} E_0(x_{1L}, x_{1R})~,
   \nonumber\\
   F^{(e)}_2 (x_{2L}, x_{\mu L}) \!\!\!&=&\!\!\!
   -\frac{1}{2}D_1(x_{2L}, x_{\mu L}) - \frac{1}{4}D_0(x_{2L}, x_{\mu L})~,  
   \nonumber\\
   F^{(d)}_1 (y_{1L}, y_{\mu L}, y_{1R}, y_{\mu R})  \!\!\! &=& \!\!\!
     \frac{y_{1R}}{54\, y_{1L}} E_0(y_{1L}, y_{1R}) 
   - \frac{1}{36} D_0(y_{1L}, y_{\mu L})  
   - \frac{y_{1R}^2}{18\, y_{1L}^2} D_0(y_{1R}, y_{\mu R})~,
   \nonumber\\
   F^{(d)}_2 (y_{2L}, y_{\mu L})  \!\!\! &=& \!\!\!
    - \frac{1}{2}D_1(y_{2L}, y_{\mu L}) + \frac{1}{4} D_0(y_{2L}, y_{\mu L})~,
   \nonumber\\ 
   F^{(d)}_3 (y_{3L}, y_{3R})  \!\!\! &=& \!\!\!
   -\frac{4\, y_{3R}}{9\, y_{3L}} E_0(y_{3L}, y_{3R})~,
   \nonumber\\
   G^{(d)}_1 (y_{1L}, y_{\mu L}, y_{1R}, y_{\mu R})  \!\!\! &=& \!\!\! 
    - \frac{y_{1R}}{18\, y_{1L}} E_0(x_{1L}, x_{1R}) 
   + \frac{1}{12} D_0(x_{1L}, y_{\mu L})  
   + \frac{y_{1R}^2}{6\, y_{1L}^2} D_0(x_{1R}, y_{\mu R})~,
   \nonumber\\
   G^{(d)}_2 (y_{2L}, y_{\mu L}) \!\!\! &=& \!\!\!
    \frac{3}{4}D_0(y_{2L}, y_{\mu L})~,
   \nonumber\\
   G^{(d)}_3 (y_{3L}, y_{3R}) \!\!\! &=& \!\!\!
   -\frac{y_{3R}}{6\, x_{3L}} E_0(y_{3L}, y_{3R}) + \frac{3\, y_{3R}}{2\, y_{3L}}E_1(y_{3L}, y_{3R})~,
\end{eqnarray}
Here, we define 
\begin{eqnarray}
 D_i(x,y) = \frac{f_i(x)-f_i(y)}{x-y}~, \quad
 E_i(x,y) = \frac{ x f_i(x) - y f_i(y)}{x-y}~,
\end{eqnarray}
for $i=1,2$ and
\begin{eqnarray}
 f_0(x)  = \frac{1-x^2+2 x \log x}{(1-x)^3}~, \quad
 f_1(x)  = \frac{3-4x+x^2+2 \log x}{(1-x)^3}~.
\end{eqnarray}
In the limit of degenerate arguments, 
we obtain $D_0(1,1) = -1/6$, $D_1(1,1) = 1/2$, $E_0(1,1)=1/6$ and $E_1(1,1)=-1/6$.

The following functions are used to calculate threshold corrections to electron and down quark masses, 
\begin{eqnarray}
  I_1^{(e)}(x_{1L},x_{\mu L},x_{1R},x_{\mu R}) \!\!\!&=&\!\!\!
   - \frac{4\, x_{1R}}{3\,  x_{1 L}} H_2(x_{1R},x_{\mu R})
   + \frac{2}{3} H_2(x_{1L}, x_{\mu L} )
    + \frac{4}{3} H_2(x_{1L}, x_{1L}/x_{1R})~,
  \nonumber\\
  I_2^{(e)}(x_{2L},x_{\mu L}) \!\!\!&=&\!\!\! 
  - 2 H_2(x_{2L},x_{\mu L})~, 
  \nonumber\\
  I_1^{(d)}(y_{1L},y_{\mu L},y_{1R},y_{\mu R}) \!\!\!&=&\!\!\!
   \frac{ y_{1R}}{4\, y_{1 L}} H_2(y_{1R},y_{\mu R})
    + \frac{1}{8} H_2(y_{1L}, y_{\mu L} )
    + \frac{1}{12} H_2(y_{1L}, y_{1L}/y_{1R})~,
  \nonumber\\  
  I_2^{(d)}(y_{2L},y_{\mu L}) \!\!\!&=&\!\!\!
   \frac{9}{8} H_2(y_{2L},y_{\mu L})~,
  \nonumber\\
  I_3^{(d)}(y_{3L},y_{3R}) \!\!\!&=&\!\!\!
   -2 H_2(y_{3L}, y_{3L}/y_{3R})~,
\end{eqnarray}
where
\begin{eqnarray}
 H_2(x,y)=  \frac{x\log x}{(1-x)(x-y)} + \frac{y\log y}{(y-1)(x-y)}~,
\end{eqnarray}
and the function becomes $H_2(1,1) =-1/2$ for the degenerate arguments.

\end{document}